\documentclass[sn-standardnature]{sn-jnl}

\usepackage{subcaption} 
\usepackage{xcolor}

\jyear{2021}%

\def\apj{Astrophys.~J.}                 
\def\aap{Astron.~Astrophys}                
\def\araa{Annual Rev.~in Astron.~Astrophys}                
\def\mnras{Mon. Not. R. Astron. Soc.}             
\def\nat{Nature}              

\newcommand{\new}[1]{\color{black}{#1}}
\usepackage[mathscr]{euscript}
\begin{document}

\title[A unifying feature for radio-emitting neutron stars]{Quasi-periodic sub-pulse structure as a unifying feature for radio-emitting neutron stars}

\author*[1,2]{\fnm{Michael} \sur{Kramer}}\email{mkramer@mpifr-bonn.mpg.de}
\equalcont{These authors contributed equally to this work.}

\author*[1]{\fnm{Kuo} \sur{Liu}} \email{kliu@mpifr-bonn.mpg.de}
\equalcont{These authors contributed equally to this work.}

\author[1]{\fnm{Gregory} \sur{Desvignes}}
\author[1]{\fnm{Ramesh} \sur{Karuppusamy}}
\author[2]{\fnm{Ben W.} \sur{Stappers}}

\affil[1]{\orgname{Max-Planck-Institut f{\"u}r Radioastronomie},
  \orgaddress{\street{Auf dem H\"ugel 69}, \city{Bonn}, \postcode{53121}, \country{Germany}}}

\affil[2]{\orgdiv{Jodrell Bank Centre for Astrophysics}, \orgname{The University of Manchester}, \orgaddress{\street{Oxford Road}, \city{Manchester}, \postcode{M13 9PL}, \country{United Kingdom}}}




\abstract{ 
  Magnetars are highly-magnetised 
  rotating neutron stars that are predominantly observed as high-energy sources. 
  Six of this class of neutron star are known to also emit radio emission, and 
  magnetars are, thus, a favoured model for the origin for at least some of the Fast Radio Bursts (FRBs).
 If magnetars, or neutron stars in general, are indeed responsible, 
  sharp empirical constraints on the mechanism producing radio emission are required. 
  Here we report on the detection of polarised quasi-periodic sub-structure in the emission of all well-studied radio-detected magnetars.
  A correlation previously seen, relating sub-structure in pulsed emission of
 radio emitting neutron stars to their rotational period, is extended, and shown to now span more than six 
 of orders of magnitude in pulse period. 
 This behaviour is not only seen in magnetars but in members of
  all classes of radio-emitting rotating 
  neutron stars, regardless of their evolutionary history, their power source 
  or their inferred magnetic field strength. 
  If magnetars are responsible for FRBs, it supports
  the idea of being able to infer underlying periods from sub-burst timescales in FRBs.
  }


\maketitle



Neutron stars manifest themselves in a number of classes. 
Arguably, the most extensively studied one is that of Galactic rotation-powered radio pulsars, 
with their emission properties investigated across the electromagnetic spectrum  \cite{lk04,pk22}. ``Normal'' pulsars have an average rotation period of about 0.6\,s, 
but some of those range from a few tens of milliseconds after birth
to a few seconds or up to 23\,s for old pulsars \cite{tbc+18}, or possibly even 76-s \cite{Caleb2022}. 
The ''millisecond pulsars'' have periods of a few milliseconds, 
obtained after a spin-up phase via mass accretion from a binary companion, 
which ``recycles'' a previously ``dead'' radio 
pulsar to enable it to become radio emitting again. 

Among the most energetic neutron stars is the class called ``magnetars'', 
neutron stars with typical rotation periods of 1 - 12
s. They emit high energy outbursts powered by their extremely large ($\sim 10^{15}$\,G) magnetic
fields \cite{kb17}, which can trigger transient radio emission as seen in six magnetars so far \cite{crh+06,efk+13,kb17}. 

Recently, interest in magnetars and their properties heightened further by
their possible connection to ``Fast Radio Bursts'' (FRBs),
which are millisecond-long bursts of radio emission
from extra-galactic sources \cite{lbm+07,tsb+13}. The
origin of FRBs is not
yet understood, but the models discussed are apparently able to explain
certain observed FRB features, such as spectra or characteristic
frequency sweeps \citep{mml19,lyu21}. Some differences in the
emission properties have been identified between signals from FRBs
that are observed to repeat \cite{sch+14} and those, where no repeating
signal has been detected so far \cite{chimecat}.  
Currently about 24 FRBs, or about
5\% of the detected FRBs, are known to have emitted more than one burst \cite{frbstatS}. It is not clear 
whether all non-repeating FRBs will
eventually be seen to repeat \cite{csrf19,jof+20}.

While the verdict on the existence of (at least two) distinct FRB 
source populations is still out, the origin of repeating FRB signals 
is clearly associated with non-cataclysmic processes. Soon after the
discovery of FRBs, magnetars were speculated to be a possible source
of FRBs due to their energetic nature \cite{pww+19}.
The recent outburst of the Galactic magnetar, SGR J1935$+$2154, showed
some FRB-like properties \cite{brb+20,chime1935,ksj+21}, 
lending credence to such
an association, although questions, for instance  whether magnetar
{\em radio} emission is luminous enough to explain FRBs, still remain (e.g.~\cite{bel21}).

If FRBs are indeed associated with magnetars, one may also expect 
to detect periodicities in repeating FRBs that are related to the 
rotation of the underlying neutron star, i.e.~in the range of a few seconds to 
tens or, possibly, hundreds of seconds. 
However, an attempt to find a periodicity between 1\,ms and 1000\,s, for instance, in hundreds of bursts from FRB 20121102A has not been successful \cite{dwz+21}.
In contrast, on sub-burst timescales, quasi-periodic sub-structure was reported \cite{mpp+21,abb+22,pvb+22} for
several bursts from FRBs, 
 though we note that
the significance of some detections has been questioned \cite{pvb+22}.
Such FRB sub-burst structure has been likened  
to quasi-periodic ``micropulses'' seen in radio pulsars (e.g.~Ref.~\cite{mpp+21}).

The short-duration micropulses in radio pulsars have a typical width, $\tau_\mu$,
and often appear quasi-periodically with a
quasi-periodicity, $P_\mu$. They are often superimposed on top of wider sub-pulses comprising the individual pulses for each rotation\cite{lk04}. It has been long established for normal, non-recycled pulsars that both
$\tau_\mu$ and $P_\mu$ scale with the rotational period,
$P_\mu \sim 10^{-3}P$, and $\tau_\mu \sim 0.5 P_\mu$
\cite{tmh75,cor79a,kjv02}. Recently, it has also been found that micropulses (also called ``microstructure'')
can be observed in recycled millisecond
pulsars \cite{dgs16,lab+22} at very short periods, which follow the same trend. 
Hence, while the names ``microstructure'' or ``micropulses'' 
were coined after discovering this emission feature in observations of
pulsars with slow spin periods (in the approximate range of 0.2-3 s)
 \cite{han71,cor79a}, in light of an increasing
range of timescales (see below), we suggest that it is more appropriate to refer to it
as ``quasi-periodic sub-structure'' that is linked to the
pulsar rotation period.

In this work, we show that this quasi-periodic sub-structure is detectable in members of
all sub-classes of radio emitting neutron stars, especially in radio-loud magnetars,
and that the previously observed dependence of sub-structure quasi-periodicity on rotational period is still obeyed. This may support
attempts \cite{mpp+21,lab+22,pvb+22} to potentially infer underlying rotational periods from
FRBs, if they were indeed emitted by extra-galactic magnetars. More importantly,
however, the observation provides a unifying feature that links the radio emission
of members in all sub-classes of radio emitting neutron stars.

\section*{Data}
To perform this study, we collected data from all radio-loud magnetars known so far: XTE J1810$-$197, Swift J1818.0$-$1607, PSR J1622$-$4950, 1E 1547.0$-$5408 (also known as PSR J1550$-$5418), PSR J1745$-$2900 and SGR J1935$+$2154. In addition, we included the recently discovered 
source GLEAM$-$X J162759.5$-$523504.3, which has also been considered
as an ultra-long period magnetar \cite{hzb+22}. 
Due to the typically flat flux density spectrum of radio-loud magnetars, our study has been conducted using data obtained at relatively high radio frequencies (i.e.~typically several GHz), with the exception of GLEAM$-$X J162759.5$-$523504.3, which was only seen at frequencies below 230 MHz.
For XTE J1810$-$197 and Swift
J1818.0$-$1607, we conducted new polarisation observations of individual pulses using the Effelsberg
100-m radio telescope (Methods). 
for each of the two sources, the
observations took place 
at 6\,GHz on three different epochs (see Table~\ref{tab:obs}). The data were recorded as a filterbank in all
four Stokes parameters, with a total bandwidth of 4\,GHz and time and
frequency resolution of 131\,$\mu$s and 0.976\,MHz, respectively. We removed the frequency dependent time delay caused by free electrons along the line of sight and formed a series of pulses which correspond to each individual rotation of the magnetar (i.e., single-pulse data). The single-pulse data were then calibrated for polarisation and flux density and cleaned to remove radio interference. For the rest of the sources, the data were collected from either public archives or published literature and analysed
accordingly. See Methods for more details.

\section*{Results}
\label{sec:results}

The best studied radio-detected magnetar is
XTE J1810$-$197 \cite{crh+06}. Quasi-periodic sub-structure in its pulses was suggested during its first phase \cite{ssw+09} and more recent phase \cite{Maan2019,crd+22}
of radio emission. Even though neither work reported a value for the quasi-periodicity,
our result appears to be fully consistent with the previous qualitative
discussion.

With this work, we have demonstrated the existence of quasi-periodic substructure in all radio-loud magnetars (see Figure~\ref{fig:pulses}), except for SGR J1935+2154 for which there are too few detected pulses with sufficient time resolution so far to make a clear detection (see Methods). 
{\new We note, however, that very recently fine-structure was
reported in sub-pulses on time scales of 5 ms \cite{zhu23}.}

The ability to detect quasi-periodic sub-structure depends crucially on the available time
resolution and the signal-to-noise ratio, while the strength of the source
can vary with time, especially for magnetars (e.g.~Ref.~\cite{ljk+08}). But even under similar
observing conditions, quasi-periodic sub-pulses are not ubiquitous and are not detectable all the time.
For instance, the magnetar PSR
J1622$-$4950 shows  such structure very prominently in observations in
2017, but less clearly before or later in 2018 (see Methods). This is very similar
to what is also observed in normal pulsars \cite{lkwj98}. Therefore, it is possible
that SGR J1935$+$2154 may reveal quasi-periodic  sub-structure
in a later, larger samples of pulses {\new (see e.g.~\cite{zhu23})}.

Examples of the detected quasi-periodic 
sub-structure
from four of the sources, XTE~J1810$-$197, Swift J1818.0$-$1607, J1622$-$4950, and GLEAM$-$X J162759.5$-$523504.3 are displayed in Figure~\ref{fig:pulses}. We measure the quasi-periodicities ($P_\mu$) and widths ($\tau_\mu$) and list
them in Table~\ref{tab:results}, which also summarises values compiled from the literature (see Methods).
Similar to normal pulsars \cite{lkwj98} and millisecond pulsars
\cite{lab+22} (Table~\ref{tab:psrvals}), when quasi-periodic sub-structure is detected
(see Methods), 
we measure a range of values forming a source-specific 
distribution (with one or two clearly preferred values) for the typical 
quasi-periodicities and widths measured 
(see e.g.~Supplementary Information).
Using the geometrical mean as a robust measure, the values of quasi-periodicities 
measured for magnetars are shown as red symbols
in Figures~\ref{fig:correlation},
\ref{fig:correlation3d} \&
\ref{fig:correlation2}. Strikingly, magnetars do not only
show  quasi-periodic sub-structure, but its quasi-periodicity also
follows the exact same scaling with rotational period that was
established for normal pulsars and millisecond pulsars
(Figure~\ref{fig:correlation}). This even extends to the
recently discovered 76-s PSR\,J0901$-$4046, where pulses are also
observed to show a set of (otherwise well-defined) quasi-periodicities
in single pulses \cite{Caleb2022}. As pointed out, the source
establishes the existence of ultra-long period
neutron stars and also suggests a possible connection to magnetars \cite{Caleb2022}.

One can even go further by considering the, also recently discovered, pulsating radio source GLEAM-X~J$162759.5-523504.3$ with a period of $1090$\,s, which is also speculated to be 
an ultra-long period neutron star or, specifically, an old magnetar \cite{hzb+22}. 
We have analysed
the available data and also identified quasi-periodic sub-structure in its pulses
(see Methods). But,
just as discussed for magnetars and normal pulsars, it is only detectable
in a limited range of epochs. The measured quasi-periodicity is somewhat
larger than a $P_\mu \sim 10^{-3}P$ scaling would suggest, i.e., we measure a value of
$6.6\pm0.6$\,s.  While we discuss this further below, the measurement nevertheless fits
the general trend, extending the observed relationship to about six orders of magnitude. 
Fitting a power-law to the whole range of sources, we obtain
$P_\mu {\rm (ms)} = (0.94\pm0.04) \times P{\rm (s)}^{0.97\pm0.05}$, 
confirming the previously inferred
scaling of $P_\mu=10^{-3}P$ \cite{tmh75,cor79a,kjv+10} determined for normal pulsars alone.

Very interestingly, the recent discovery of a 
Rotating Radio Transient (RRAT), RRAT~J1918$-$0449, 
with the Five Hundred Meter Aperture Spherical Telescope, 
allows to resolve and for the first time detect 
quasi-periodic sub-structure in pulses from RRATs
\cite{cwy+22}. The measured quasi-periodicity agrees perfectly with our scaling law (see Figure~\ref{fig:correlation}).

The measurements of the sub-structure width, $\tau_\mu$, are sometimes more difficult 
to obtain than those of quasi-periodicities, as pointed out for normal pulsars \cite{mar15}. 
This results in fewer measurements or larger uncertainties,
but we also find a linear scaling with rotational period as shown 
in Figure~\ref{fig:correlation2}  and
express this fit to the data as 
$\tau_\mu {\rm (ms)} = (0.59\pm0.03) \times P{\rm (s)}^{0.99\pm0.02}$. This is consistent with previous findings from normal pulsars alone, but now also expanded in period range and,
in particular, in type of radio-emitting neutron star.

Given that both independently measured quantities, $\tau_\mu$ and $P_\mu$, 
obviously depend on rotational period, $P$,
we also perform a joint fit for those sources, where both measurements can be made.
We determine the joint power law index, $\alpha$, and the independent scale factors,
$A_P$ and $A_\tau$, i.e.
$P_\mu$(ms)$= A_P \times \; P$(s)$^\alpha$ and
$\tau_\mu$(ms)$ = A_\tau \times P$(s)$^\alpha$. We again confirm a linear dependency
on rotational period, $\alpha = 1.03\pm 0.04$, while the spacing of the 
quasi-periodic sub-pulse structure
is approximately twice their width, i.e.~$A_P = 1.12 \pm 0.14$ and $A_\tau=0.50\pm0.06$. The results 
are shown in Figure~\ref{fig:correlation3d}.

\section*{Discussion}
\label{sec:discussion}

We have established an universal relationship between 
the rotational period and emission features that can be found in members 
of {\em every type} of radio-emitting neutron star. 
The relationship scales simply with rotational period and applies regardless of the formation or evolutionary history of the neutron star, 
its presumed energy source, or the regularity (or sporadicity) of its radio emission. 

The relationship extends over about six orders of magnitude. But some deviations are observed.
Firstly, the quasi-periodic sub-structure seen in magnetar pulses is not always present and, secondly,
the periodicities vary to some extent, leading to a classification as a quasi-periodicity within a pulse and  
leading to a narrow distribution of 
quasi-periods from pulse-to-pulse that is nevertheless specific for a given source.
But this is also well known for many normal pulsars and was also recently established
for the 76-s PSR\,J0901$-$4046, and shown here for GLEAM$-$X J162759.5$-$523504.3. 
In other words, the similarity goes beyond the 
existence of the scaling relationship itself. Future measurements will help to establish and constrain the
range of values further to identify more commonality.

The scatter in the measurements around the predicted values is clearly caused to some extent by the
observed fluctuation in periodicities around preferred values. We can speculate whether the measured magnetar 
periodicity values really tend to be somewhat physically larger than those values for normal pulsars, as
suggested by Figure~\ref{fig:correlation}, but the
deviations by themselves are hardly statistically significant given the uncertainties (see Methods). 
Similar thoughts should
also apply to the measurements for GLEAM-X~J$162759.5-523504.3$. But here, we also note
that unlike for the other magnetars studied here, the observations
of GLEAM-X~J$162759.5-523504.3$ were made at a rather low frequency of $\lesssim 230$ MHz.

Even though pulsar sub-structure properties are usually consistent across large frequency ranges (e.g.~\cite{cor79a,lkwj98}), observations at lower frequencies could possibly lead to broader sub-structures and larger spacing compared to the much  higher radio frequencies 
used here otherwise (see e.g.~\cite{kjv02}).
At the moment, the limited range of available epochs prevents further studies.  

The origin of the quasi-periodic sub-structure in the radio emission has been
a matter of debate since it was first discovered in normal pulsars. It
was interpreted either as a temporal or angular phenomenon \cite{cor79a}. A temporal phenomenon
would suggest emission patterns which originate ultimately from
processes in the interior or on the surface of neutron stars (e.g.~\cite{cr04c}). In contrast, interpreting
the observations as an angular pattern, the 
 quasi-periodic sub-structure represents ``beamlets'' of a characteristic
angular width that sweep across the observer with the pulsar rotation \cite{cor79a}. 
It has already been argued for normal pulsars \cite{kjv02} that a $P_\mu-P$-correlation 
is more naturally explained as an angular pattern. Given the extent of the relationship in period space
and to members of all types of radio-emitting neutron stars, 
an angular beamlet interpretation of the quasi-periodic sub-structure
appears as the only viable explanation. Interestingly,
recent work \cite{tom21a,tom21b} appears to derive the exact relationship that we see in Figure~\ref{fig:correlation}.
Applying their experience with Tokamak fusion experiments, it connects radio emission of 
rotating neutron stars to slow tearing instabilities feeding off an inhomogeneous twist 
profile within the open pulsar circuit \cite{tom21a}. In this picture, radio emission occurs in the form of 
coherent curvature emission created by Cerenkov-like instabilities of 
current-carrying Alfven waves in thin current
sheets with relativistic particle flow \cite{tom21b}.
The model, which is also applicable to magnetars, predicts that beamed
radio emission is created from packets of charged particles, with 
 quasi-periodic sub-structure being a natural consequence.
The encountered timescales are constrained by relativistic beaming, deriving a
scaling relationship where the  relevant timescale 
depends only on the angular frequency $\Omega$ and an effective gamma-factor, $\gamma_{\rm eff}$, i.e.
$t_\mu \sim 1/\Omega\gamma_{\rm eff}$, predicting a period dependency 
of the order of $10^{-3} P$ \cite{tom21b}, as we observe. Even though
a range of size structures may be present in the current profile in the pulsar circuit, possibly
explaining the fluctuation in the observed quasi-periodicities of a particular source,
only the largest are visible due to the smearing effect of relativistic emission \cite{tom21a,tom21b}.
The universality of the scaling law shown here suggests that the effective gamma-factor
may be the same for all radio emitting neutron stars, i.e.~ $\gamma_{\rm eff}\sim 200$.

The fundamental dependence of emission sub-structure (i.e. quasi-periodicities and also pulse width, see
Figures \ref{fig:correlation}, \ref{fig:correlation3d} \& \ref{fig:correlation2}) on rotational period among all 
types of radio-loud rotating
neutron stars is especially intriguing for the magnetic-field powered 
magnetars, even though young (rotation powered)
pulsars and magnetars have been
known to show some similarities in their emission features
\cite{ksj+07,pk22}; Radio pulsars usually have a steep flux density spectrum \cite{lk04}, while the spectrum
of younger pulsars tend to be flatter \cite{jvk+18} and that of magnetars is usually very
flat or even inverted \cite{ljk+08,torne22}
or may also be complex sometimes \cite{Lower2020}.

In contrast, the polarisation features of magnetars and young and energetic pulsars (defined here as those with
 $\dot{E}>10^{35}$ erg s$^{-1}$) are very similar. 
 The latter often show a very large degree of linearly polarised emission \citep{wj08}, 
  while magnetars are also typically 100\% linearly polarised \cite{ksj+07,pk22}. The 
corresponding position angle (PA) of the average pulse profile of pulsars usually shows 
a distinct variation as a function of pulse duration \citep{wj08}, but it tends to be flatter
or irregular for magnetars \cite{ksj+07,pk22}. 
However, quasi-periodic sub-structure from {\em individual} rotations in normal
pulsars can also exhibit an apparently flat PA swing
\cite{mar15}. Our observations (see Figure \ref{fig:polpulse}) clearly demonstrate that this
is also the case for polarised quasi-periodic sub-structure emission in magnetars.

It has been noted before that these magnetar
polarisation features are akin to those of bursts from repeating FRBs \cite{hsm+21}, although exceptions exist \cite{Luo2020,ksl+21}. 
If magnetars are indeed responsible for (some or all) FRBs, suggestions 
that quasi-periodic sub-structures in FRBs may be similar to that in normal pulsars
\cite{phl19,abb+22,lab+22} would imply that similar plasma processes 
(and not vibrations of the neutron star  \cite{mvh88}) are responsible.
If periodicities abide by the same scaling law, one can follow
suggestions \cite{mpp+21,lab+22,pvb+22} to use the observed timescales for deriving the underlying rotation period,
as also demonstrated in Fig.~\ref{fig:correlation}. In this case, 
one may wonder why underlying rotation periods have not yet been detected, e.g.~in detailed 
studies of FRB repeaters such as Ref.~\cite{dwz+21}. However, the ability to find a periodicity
can be hampered severely if the pulse window were to extend over a 
significant fraction of the rotational period \cite{lbh+15}. 
Indeed, magnetar radio emission tends to have much wider duty cycles than normal pulsars \cite{pk22}. 
For instance, PSR J1622$-$4950 emits across almost the full rotation  \cite{lbb+10}.
A large pulse duty cycle may be caused by a non-dipolar magnetic field structure in
magnetars, and implies
a generally much wider beam than for normal pulsars, covering a much wider area of sky. 
If that were also the case for FRBs, it would reduce the number of FRBs inferred from population studies accordingly.

In summary, we have demonstrated the existence of quasi-periodic sub-structure in the radio
emission of members of all types of radio-emitting rotating
neutron stars, regardless of their evolutionary history, their power source or their
inferred
magnetic field strength. Whether magnetars are related to FRBs remains to be seen, but 
we have shown that quasi-periodic emission structures exist in the individual pulses
of radio-loud magnetars, often showing a high degree of polarisation with flat position angles,
that follows the same dependence on rotational periods as found for normal and millisecond pulsars.
This universal relationship now spans six-orders of magnitude in rotational
period. Considering the various types of radio-emitting neutrons stars as whole,
we obtain tantalising insight that all appear to
share some similar fundamental processes in their magnetosphere.

\clearpage

\section*{Methods}
\label{sec:methods}

\setcounter{figure}{0}
\setcounter{table}{0}

\subsection*{Observations and data collection} 

The data utilized for this study result from new observations, re-analysed
archival data and published results, covering all six radio-detected magnetars known to date. New observations of two magnetars, XTE J1810$-$197 \cite{crh+06} and Swift J1818.0$-$1607 \cite{ccc+20}, were carried out each on three epochs using the Effelsberg 100-m telescope and its CX-band receiver which covers observing frequencies of 4--8\,GHz. 
Data acquisition was made with a pulsar backend consisting of two units of the second generation of the Reconfigurable Open Architecture Computing Hardware developed with the Field Programmable Gate Array technique by the CASPER team\cite{casper}. The data were recorded in 8-bit samples stored in \textsc{psrfits} search mode format \cite{hsm04}, with a time resolution of 131\,$\mu$s and 4096 frequency channels. Single pulse
data  with specifications detailed in Table~\ref{tab:obs} were extracted from the search-mode data stream; this used the \textsc{psrfits\_utils} software toolkit (see {\tt https://github.com/demorest/psrfits\_utils}) and an ephemeris obtained from our regular timing program on these two sources. Dispersion measures (DMs) of 178.0 and 703.0\,cm$^{-3}$\,pc were used to de-disperse the data of XTE J1810$-$197 and Swift J1818.0$-$1607, respectively \cite{crh+16,ccc+20}. Next, the single-pulse data were calibrated for polarisation, flux density and cleaned for radio interference using the \textsc{psrchive} software package \cite{hsm04}. Polarisation calibration
was obtained using information from an injected noise diode signal associated with the pulsar observation. The 
data were flux-density calibrated using an exposure on the calibration source NGC~7027 and its catalogued
flux density spectrum
\cite{zvp08}. Finally, rotation measures (RMs) of 74.44 and 1442\,rad\,m$^{-2}$ were applied to the data of XTE J1810$-$197 \cite{crh+06} and Swift J1818.0$-$1607 \cite{crh+16,ccc+20}, respectively, to correct for 
interstellar Faraday rotation in the linear polarisation component.

Data of magnetar J1622$-$4950 measured at 3.1\,GHz with the Parkes telescope \cite{lbb+10} were 
obtained from the public ATNF data archive (see {\tt https://atoa.atnf.csiro.au}); 
the selection of data focused on epochs when the magnetar was exceedingly bright allowing for detailed single-pulse analyses. These data are available in \textsc{psrfits} search format with total intensity information and
specifications detailed in Table~\ref{tab:obs}. They were  processed to generate single-pulse data with the \textsc{dspsr} software package, an ephemeris obtained from Ref.~\cite{scs+17} and a DM of 820\,cm$^{-3}$\,pc. 

Pulsation data on the magnetar-like periodic radio transient GLEAM$-$X J162759.5$-$523504.3 were obtained from the data archive published by \cite{hzb+22}. The observations were carried out with the Murchison Widefield Array at
 88 MHz to 215 MHz. 
In addition, we obtained information on pulse properties of magnetar 1E 1547.0$-$5408 from data retrieved 
from the ATNF archive, and collected information from
the literature on the well-studied Galactic Centre magnetar, PSR J1745$-$2900, and the sparsely detected  
SGR J1935+2154 (see the next section for details).

\subsection*{Measurement of magnetar sub-structure's widths and quasi-periodicities} \label{ssec:ACF_measure}
{\it XTE J1810$-$197, Swift J1818.0$-$1607, PSR J1622$-$4950 and GLEAM$-$X J162759.5$-$523504.3}:
For these three magnetars and the GLEAM source,
the sub-structure's quasi-periodicities and widths were measured using the auto-correlation analysis. The methodology of this analysis is well summarised in Section 7.4.2.2 of Ref.~\cite{lk04} and many other publications \cite[e.g.,][]{tmh75,cor76a}. Here, the first principle of this methodology is also shown in the sketch in the Supplementary Information. For a given waveform $I(n)$, its auto-correlation function (ACF) is defined as:
\begin{equation}
    \textit{A}(k) = \sum_n I(n+k)I(n).
\end{equation}
When there is quasi-periodic structure in the pulse, it manifests itself by exhibiting a sequence of equally spaced local maxima in the ACF. The time lag of the first local maximum corresponds to the characteristic separation, i.e., quasi-periodicity of the
sub-structure, while those of the rest equal to incremental numbers times the periodicity (see Supplementary Information). Thus, the value of the characteristic quasi-periodicity  can be defined as the first local maximum in the ACF. In order to search for quasi-periodic sub-structure and to determine a value for the periodicity, we first calculated the ACFs for all of the single pulses in our data. Then we used a Fourier Transform to calculate the Power Spectral Density of each of the pulse and its ACF, and in both identified a set of maxima each of which may correspond to the presence of a periodicity. Next we cross-checked the two groups of maxima and kept those which were reported in both (within 20\% difference in values) as candidates of the periodicity. Finally, we visually checked the pulse profile and its ACF, and noted the pulse as a detection only when at least one of the reported periodicity corresponds to a real periodic feature in both the profile and its ACF. The shortest reported candidate periodicity was recorded as the value of the detected quasi-periodicity in the pulse. In total, 452, 4126, 208, 12 pulses from XTE J1810$-$197, Swift J1818.0$-$1607, PSR J1622$-$4950 and GLEAM$-$X J162759.5$-$523504.3, respectively, 
were investigated and the number of pulses detected with quasi-periodic sub-structure were in turn 77, 400, 93, 2. The results
are displayed in corresponding Figures of the Supplementary Information.
We note that for XTE J1810$-$197 we observe a bimodal distribution of periodicities, as it is occasionally observed for normal pulsars \cite{mar15}. We note that our derived value is at the lower end implied by the averaged ACF shown by Ref.~\cite{crd+22}, whereas a value reported by Ref.~\cite{Maan2019} is consistent with the first peak of our distribution.

We also measured the width of the sub-structure using the ACF analysis. As in our data the inverse of the channel bandwidth is smaller than the used sampling times,  
the width can be defined as the first turning point in the ACF from zero lag (see Supplementary Information). 
Its value corresponds approximately to the Full Width at Half Maximum of the sub-structure \cite{cwh90,lkwj98}. To obtain the turning point for each pulse, a spline fit was conducted to the ACF starting from the first non-zero lag bin until the first bin before the identified quasi-periodicity, which reported a group of detected knots. Then all reported knots were visually inspected together with the ACF, to identify the exact one that corresponds to the first turning point in the ACF for each pulse. 

As observed for normal pulsars, not all individual pulses exhibit identifiable quasi-periodic sub-structure. Also similar to normal pulsars, a limited range of periodicities can sometimes be observed. This is caused by a mixture of intrinsic variation, the occurrence of harmonically related quasi-periodicities, or instrumental and observational constraints (see e.g., \cite{cor76a,lkwj98}). For this reason, 
we studied the distribution of values measured for the ensemble of studied pulses. 
In order to account for possible cases where
the histogram does not show a single, clearly identifiable peak, we refer
to the geometric mean as our preferred values as a robust method to determined a preferred value. In Table~\ref{tab:results}
we quote uncertainties of the geometric mean,
$x_{\rm geo}$, expressed in the form of the geometric standard deviation (GSTD) factor, $\Delta x_{GSTD}$, 
defining a range from $x_{\rm geo}/\Delta x_{GSTD}$ to $x_{\rm geo} \times \Delta x_{GSTD}$. We also quote
the mean, the error of the mean and the median for comparison.

\medskip

There are three more magnetars that have been detected at radio frequencies: 1E 1547.0$-$5408 (also known as PSR J1550$-$5418) \cite{crhr07}, the Galactic Centre magnetar PSR J1745$-$2900 \cite{efk+13} and SGR J1935$+$2154 \cite{brb+20,chime1935,ksj+21}. Below we discuss the obtained sub-structure properties for these sources in turn, using partly archival data and previously published results.

{\it 1E 1547.0$-$5408}: This 2.1-s magnetar showed strong transient radio emission,
especially after its 2009 outburst \cite{camilo2009}. Retrospectively, for this
outburst, two strong
radio pulses were reported recently \cite{ibr+21}, which saturated the 1-bit digitisation of the 
observing system and potentially distorted the pulse signal. Thus, we chose to study a different observation with the Parkes telescope from 2009, February 25,
(MJD 54887), $17$ UT, at 8.3\,GHz. With a time resolution of 1\,ms, this 30-min observation reveals very narrow
pulses leading the main pulse by about 500\,ms. To our knowledge, these features have not yet been reported.
The recorded sub-integrations contain 9 periods
each, but the spacing of these pulse suggests that they are the result of single bright bursts 
occurring at slightly different rotational phases during the folded 18\,s. The pulses are
consistent with having a width $\lesssim 1$\,ms. Inspecting the spacing of these pulses, one
can infer a periodicity of about 4\,ms. These estimates are consistent with an
analysis of further search-mode 
data accessible from the archive and recorded during 2016 and 2017, although the available
time resolution here is also limited to 1\,ms. Given these constraints, we consider these measurements
with caution but list them for completeness with a GSTD of 2.0.

{\it J1745$-$2900}: This magnetar exhibits the longest duration of 
uninterrupted detectable radio emission since its
first detection in the radio in 2013 \cite{efk+13}. The profile has been observed to be very variable
in frequency \cite{tek+15} and on short and long timescales \cite{wcc+19}, 
with its strength diminishing in recent years \cite{scc+21}.
The strong background emission from Sgr A* combined
with the relatively long period of 3.76\,s causes significant red noise in single-dish timeseries measurements,
and with a very large dispersion measure of $1778$ 
pc cm$^{-3}$, this implies that high-frequency observations
with interferometers are best suited to study the existence of quasi-periodic
sub-structure. Summarizing the wealth 
of observations, the broad pulse envelope often reveals partly overlapping subpulses with
clearly discernable short pulses (see e.g. Figure~3 of \cite{pwp+18} of Figure 4 of
Ref.~\cite{wcc+19}). The typical intrinsic width of individual emission components 
is found to be as short as 1.6\,ms \cite{wcc+19} or 1.8\,ms \cite{pwp+18}. Such
narrow components, when resolvable, are separated from each other, often in a quasi-periodic
fashion, with typical
separations from $2.2 \pm 0.7$ ms (see Figures 4 and 8 of \cite{wcc+19}) to about $10.4 \pm 1.7$ ms  
(see Figure 3 of \cite{pwp+18}). More often these emission features are somewhat wider and
blend into each other, overlapping at about a sub-structure width. 
The most common pulse width for all components is either 3.2 or 6.4 ms \cite{wcc+19}. This is consistent with 
recent pulse width of $4.9 \pm 0.5$ ms measured between 4.4 and 7.8\,GHz and 
corrected for interstellar scattering \cite{scc+21}. For our purposes, we account for the variety
of measurements by computing the geometrical mean from the above values, i.e.~ a pulse width of 3.1 ms (with a GSTD of 1.7, cf.~Figure 5 of \cite{wcc+19}) and a quasi-periodicity of 4.8 ms (GSTD 1.9).

{\it SGR J1935$+$2154:} Efforts to detect radio emission from this 3.24-s magnetar 
following its outbursts had failed repeatedly, providing stringent upper limits on existence
of detectable radio emission \cite{ykj+17}. In contrast,
on April 28, 2020, both the CHIME \cite{chime1935} and STARE2 \cite{stare1935} telescopes detected the
same short radio burst from a direction consistent with that of SGR J1935$+$2154, which were followed by a sequence of detections within a short time window \citep[e.g.,][]{ATEL14074,ATel14080,ATel15681}.
Both CHIME and STARE2 measured a DM around 333\,pc cm$^{-3}$.
The CHIME data observed between 400 and 800 MHz revealed two sub-bursts with 
widths determined to 0.585(14)\,ms and 0.335(7)\,ms, respectively, after
correcting for apparent effects of  multi-path scattering (with a thin-screen scattering timescale of 0.759(8)\,ms 
when referenced to 600\,MHz). Both components were separated by 28.91(2)\,ms. The STARE2 data taken
at higher frequencies between 1280\,MHz and 1530\,MHz showed no significant evidence for scattering
\cite{stare1935}. Nevertheless, because CHIME adopted a scattering model, they also fitted a corresponding
model, deriving an apparent intrinsic width of 0.61(9)\,ms and a scattering timescale of 0.4(1)\,ms when
referenced to 1\,GHz. Scaling the CHIME scattering time to the same frequency, using the expected radio frequency
dependency of $\nu^{-4}$ \cite{tsb+13}, one finds 0.098(1)\,ms, which is clearly inconsistent with the STARE2
measurement. Further detections of radio pulses to clarify this matter turned out to be difficult.
Observations of the magnetar with the FAST telescope on April 30, 2020, detected
a weak radio pulse with a DM consistent with the CHIME and STARE2 events. While this
indicated that all three radio pulses were emitted by the magnetar, a reliable width
measurement was not reported \cite{atelFASTdetect}. Further follow-up radio observations
had mixed successes, mostly reporting non-detections (see e.g.~Refs.~\cite{bbb+21,yhb+22} and references therein),
including even deep unsuccessful FAST observations \cite{lzw+20}. A later successful detection with FAST
appeared again to be too weak to study the pulse properties \cite{1935weiwei}. More recently, however,
using over 500h of follow-up observations, two further radio bursts from the magnetar
were reported \cite{ksj+21}.
These bursts showed an observed width of 0.866(43) ms and 0.961(48) ms, respectively, separated in time by 
about 1.4\,s, at a frequency of 1324\,MHz.
Attempts to fit a scattering tail to the slightly asymmetric burst shapes,
derived scattering timescales of 0.315(12) ms and 0.299(29) ms, respectively. At a reference frequency of 1\,GHz,
this would amount to an average of about 0.952(21) ms, for the expected frequency scaling. Since
the three measured timescales are barely consistent with each other, unless one postulates an unusually
flat frequency dependence, we follow the discussion in Ref.~\cite{ksj+21}.~and their
suggestion that it is more straightforward to reconcile the measurements by assuming that the slight
asymmetries observed in the bursts are actually intrinsic. As the $1/e$-timescale of
pulse smearing due to scattering can be usually accounted for, to first order, by adding the
widths in quadrature, we derive from the published value estimates for the originally observed pulse width 
as follows:  0.73(36) ms for the STARE2 detection, 0.96(48) ms and 0.93(41) ms for the two CHIME
sub-bursts, respectively. Computing the geometric mean of all five measurements,
we derive 0.88 ms and with a GSTD of 1.2. The paucity of events registered for SGR J1935$+$2154
prevents a study of possible periodicities within the magnetar bursts beyond noting 
that the two CHIME components were separated by 28.91(2) ms. 
{\new We note, however, that very recently fine-structure was
reported in sub-pulses on time scales of 5 ms \cite{zhu23}.}

\subsection*{Measurement of pulsar sub-structure widths and quasi-periodicities}

The period-scaling of pulse sub-structure timescales and periodicities was already established for
normal pulsars in the late 1970s. Since then many more measurements have 
been made, and we make an attempt to compile them here. 
Meanwhile, quasi-periodic sub-structure has been also detected
in a sample of millisecond pulsars, which we also include to expand the study
of the period-scaling to the shortest periods. In recent years, a number of pulsars 
have been discovered with periods above 10\,s and even more recently with even larger
periods, as discussed above.

{\it Normal pulsars:}
 Quasi-periodic sub-structure known as ``microstructure''
has been studied extensively for normal pulsars, measuring 
both width and quasi-periodicities. We compiled a list of values for normal 
pulsars from Refs.~\cite{tmh75, cor76a, cor79a,cwh90,lkwj98,kjv+10} and references therein.
We also used quasi-periodicity measurements presented in Ref.~\cite{mar15},
reevaluated some of their measurements for weak pulsars,
based on the data provided by the authors with the publication.
In cases where values differed across different studies, we again computed
the geometrical mean and GSTD factors as shown in Figure~\ref{fig:correlation}.

{\it Fast-spinning millisecond pulsars:}
In recent years, measurements of quasi-periodic sub-structure have also
become available for fast-rotating millisecond pulsars
using high-sensitivity and high-time-resolution observations \cite{dgs16,lab+22}.
As for normal pulsars, we do not consider measurements of giant pulses (and their
nanoshots) as they
may have a different origin compared to regular pulsar radio emission \cite{lk04}, but concentrate
on the studies of normal pulses emitted by millisecond pulsars.
The combined measurements for both the width and quasi-periodicity follow the period
scaling as pointed out by Refs.~\cite{cor79a,kjv+10,dgs16,lab+22}. 

{\it Long-period radio pulsars:}
Since magnetars share the same period range as long-period pulsars, we also include 
unusually slowly rotating pulsars, namely 
the 8.5-s PSR~J2144$-$3933, 
the 12.1-s PSR~J2251$-$3711 and the 23.5-s PSR~J0250$+$5854. Recently, 
the presence of quasi-periodic sub-structure was reported
for PSR~J2144$-$3933 \cite{mbma20}, with
an estimate of sub-structure quasi-periodicity (as a median derived from a distribution
of 355 single pulses) of $P_\mu = 11.8\pm 0.6$ ms. The width of the
sub-structure shown in Ref.~\cite{mbma20} is consistent with a value half this size,
hence we estimate $\tau_\mu = 5.9\pm 0.3$. We convert this into corresponding estimates
for the geometric means accordingly.

For J2251$-$3711, Ref.~\cite{mke+20} presents single pulses with sub-pulses with a typical width
of 4 to 16 ms. No clear quasi-periodic sub-structure is visible, but we tentatively adopt
a typical sub-structure
width of $10\pm 6$ ms, or in terms of a geometric mean of 9.8 ms with
a GSTD of 1.6.

PSR J0250$+$5854's average profile and single pulses show differences across
frequencies \cite{awb+21}. Single pulses have only been detected confidently at
at 320\,MHz, where the strong individual pulses appear only in the first pulse
component. However, no reliable detection of quasi-periodic sub-structure 
can be made from
the published data, preventing us from including it in our analysis.

{\it Rotating Radio Transients (RRATs):}
This sub-set of rotating neutron stars emits radio pulses sporadically \citep{mll+06}. 
First considered to be a different class of neutron stars, continued or
more sensitive observations often allow to associate the emission with an underlying
period which increases with time, confirming a rotational origin of the period \citep{mlk+09}.
Ref.~\cite{kkl+11} argued that RRATs are not a distinct or a separate population,
but an extreme class of ordinary pulsars, which happen to be discovered more easily via 
their single pulses. Observations of RRATs with sufficient time resolution and sensitivity
to reveal potential quasi-periodic sub-structure are rare. Recently, however, the
2.48-s RRAT J1918$-$0449 was discovered and monitored
with the Five Hundred Metre Spherical Aperture Telescope (FAST)
 showing 
clear quasi-periodic sub-structure with a periodicity of $P_\mu = 2.31 \pm 0.25$ ms. 
The sub-structure width is measured to be $\tau_\mu = 1.47 \pm 0.25$ ms \cite{cwy+22}.

The results of our magnetar measurements, combined with the values for normal pulsars, millisecond pulsars and the recent discoveries of PSR\,J0901$-$4046, GLEAM-X J162759.5$-$523504.3 and RRAT J1918$-$0449 are shown in 
 Tables \ref{tab:results} and in Table~\ref{tab:psrvals}. They
 are summarised in Figures~\ref{fig:correlation}, \ref{fig:correlation3d}
and \ref{fig:correlation2}, for the sub-structure's quasi-periodicities and widths, respectively.

We have modelled the data with power-law fits.
We first conducted a normal least-square fit by minimising a standard $\chi^2$ expression. This provided initial guesses for two fitted parameters, power-law index and scale factor, and their uncertainties. In order to check for co-variances and the shape of posterior distributions, we then employed the UltraNest sampler \cite{ultranest} to perform an analysis with uniform priors, covering a range of $\pm100$ times the uncertainties around the least-squares results. As a likelihood function we used the usual 
$\ln({\cal L}) = -\chi^2/2$.
The resulting power-law fits and the posterior distributions are presented in the main text and the figure captions. 
They all have in common that both quantities depend linearly on rotational period of the neutron star.

With the quasi-periodicities of magnetars somewhat larger than the values implied by the scaling law, we performed a weighted least-squares fit to the magnetar data points alone. We assumed the same linear scaling with rotational period, $P$, as in Figure \ref{fig:correlation3d}  to determine only the scaling factors.  We derive for the periodicity: 
$A_P^M = 1.69\pm 0.17$  which compares to $A_P = 1.12\pm 0.14$ for the whole data set.
This is a factor of $1.51 \pm 0.24$ larger. A similar fit for the magnetar sub-structure widths finds $A_\tau^M =0.37 \pm 0.10$, which compares to 
$A_\tau = 0.50\pm0.06$, i.e. being a factor $0.74 \pm 22$ smaller. Both scalings are consistent within 2 sigma.


\subsection*{Polarisation of quasi-periodic sub-structure}

We analysed the polarisation of all quasi-periodic sub-structure from XTE~J1810$-$197 and Swift J1818.0$-$1607 obtained during our new observations with the Effelsberg Radio Telescope. The polarisation profile and its corresponding linear polarisation position angles (PA) of the pulses shown in Figure~\ref{fig:pulses} and additional examples (shown in the Supplementary Information) are presented in Figure~\ref{fig:polpulse}. For XTE~J1810$-$197, most sub-structures are close to be 100\% linearly polarised, consistent with a very high degree of linear polarisation of the integrated profile of this magnetar. The PA swings of these quasi-periodic sub-structures are apparently flat  within several percentages of the rotational period. For 
quasi-periodic sub-structures 
from Swift J1818.0$-$1607, the fraction of linear polarisation is high on average while exhibiting some distinct pulse-to-pulse variability. In some pulses the linear component can be close to 100\% while in others it may not turn out to be significant. The PA swings of most sub-structure are shown to be flat, some with low-level variation on a scale of roughly 10\,deg.  These features of the PA swings are similar to what have been seen in quasi-periodic sub-structures from ordinary pulsars \citep{mar15}.

\clearpage

\backmatter

\bmhead{Data availability}
 Most data used in this study are available from the literature or already downloaded 
 from publicly accessible archives (see {\tt https://atoa.atnf.csiro.au}). Magnetar data 
 obtained especially for this study are available by contacting
 the corresponding author to arrange the data transfer 
 due to the large data volume for these observations. \\

\bmhead{Acknowledgements}

This manuscript makes use of observations conducted with the 100-m
radio telescope in Effelsberg owned and operated by the
Max-Planck-Institut f\"ur Radioastronomy. We are grateful to Duncan Lorimer and David Champion for comments on the manuscript and thank Laura Spitler and her group for stimulating discussions. We thank Robert Main for useful discussions on microstructure in the FRBs. 
MK, KL, GD 
acknowledge the financial support by the European Research Council for the ERC Synergy Grant BlackHoleCam under contract no.~610058. BWS acknowledges the financial support by the European Research Council for the ERC Advanced Grant MeerTRAP under contract no.~694745.
\\

\bmhead{Author contributions}
 MK and KL drafted the manuscript with suggestions from co-authors. MK
 and KL reduced and analyzed the Effelberg 100-m telescope data and archival data. KL,
 GD and RK conducted observations with the 100-m radio telescope and
 preprocessed data.\\
\bmhead{Conflict of interest/Competing interests} 
The authors declare no competing interests.

\clearpage 

\section*{Figures}

\begin{figure*}[!htp]
\includegraphics[width=0.95\textwidth]{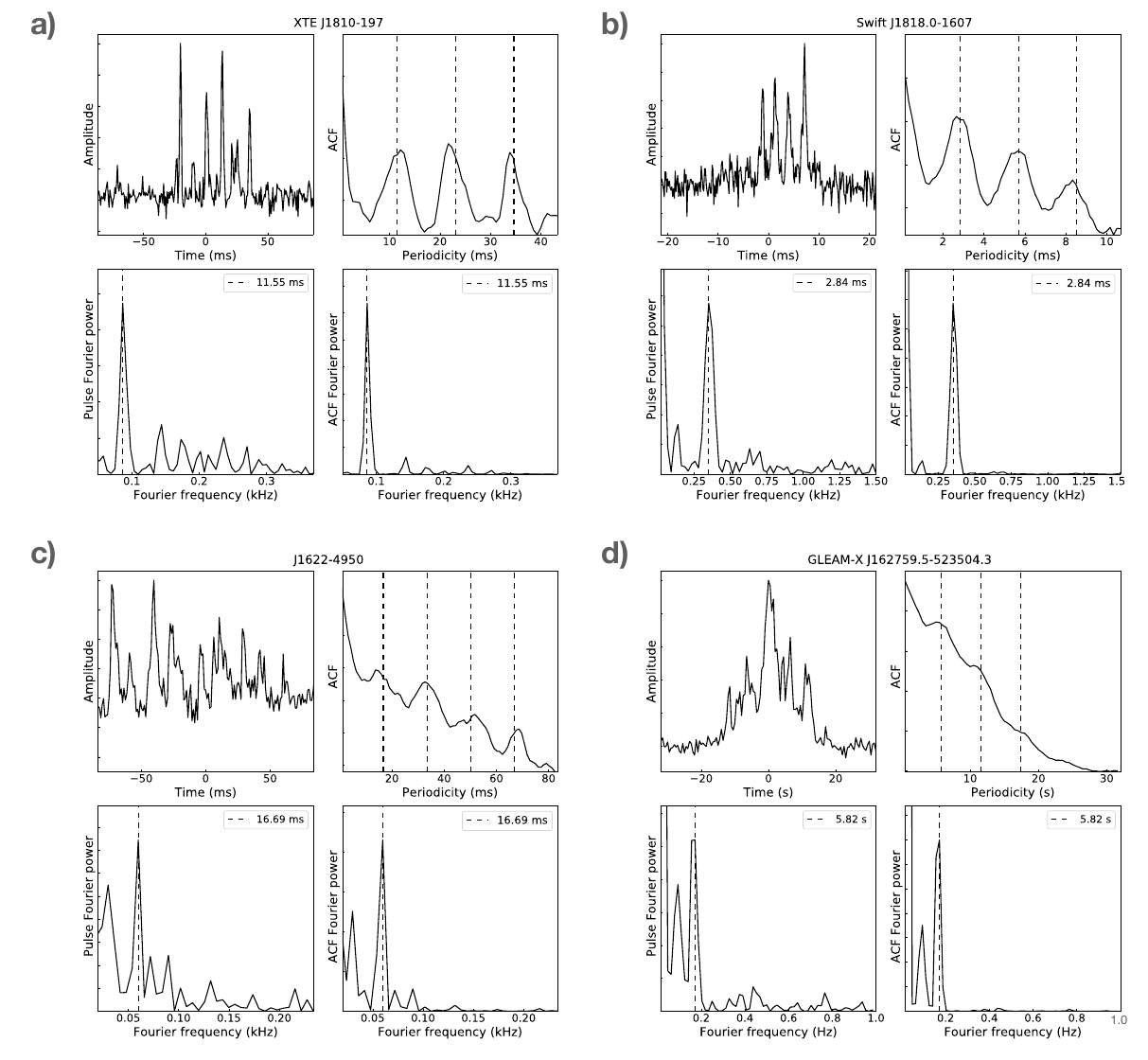}

\caption{Examples of quasi-periodic sub-structure from four of our selected sources:  a) XTE~J1810$-$197, 
b) Swift J1818.0$-$1607, c) J1622$-$4950, d) GLEAM$-$X J162759.5$-$523504.3. For each source, the four sub-panels show the pulse amplitude as a function of time, the auto-correlation (ACF) function of the pulse, the Fourier power spectral density (PSD) of the pulse, and the PSD of the ACF. The vertical dashed lines mark the identified quasi-periodicity and its harmonics in the ACF panels, and their corresponding fundamental frequency in the PSD panels.}
\label{fig:pulses}
\end{figure*}

\begin{figure}[!htp]
\centering
\includegraphics[width=\textwidth]{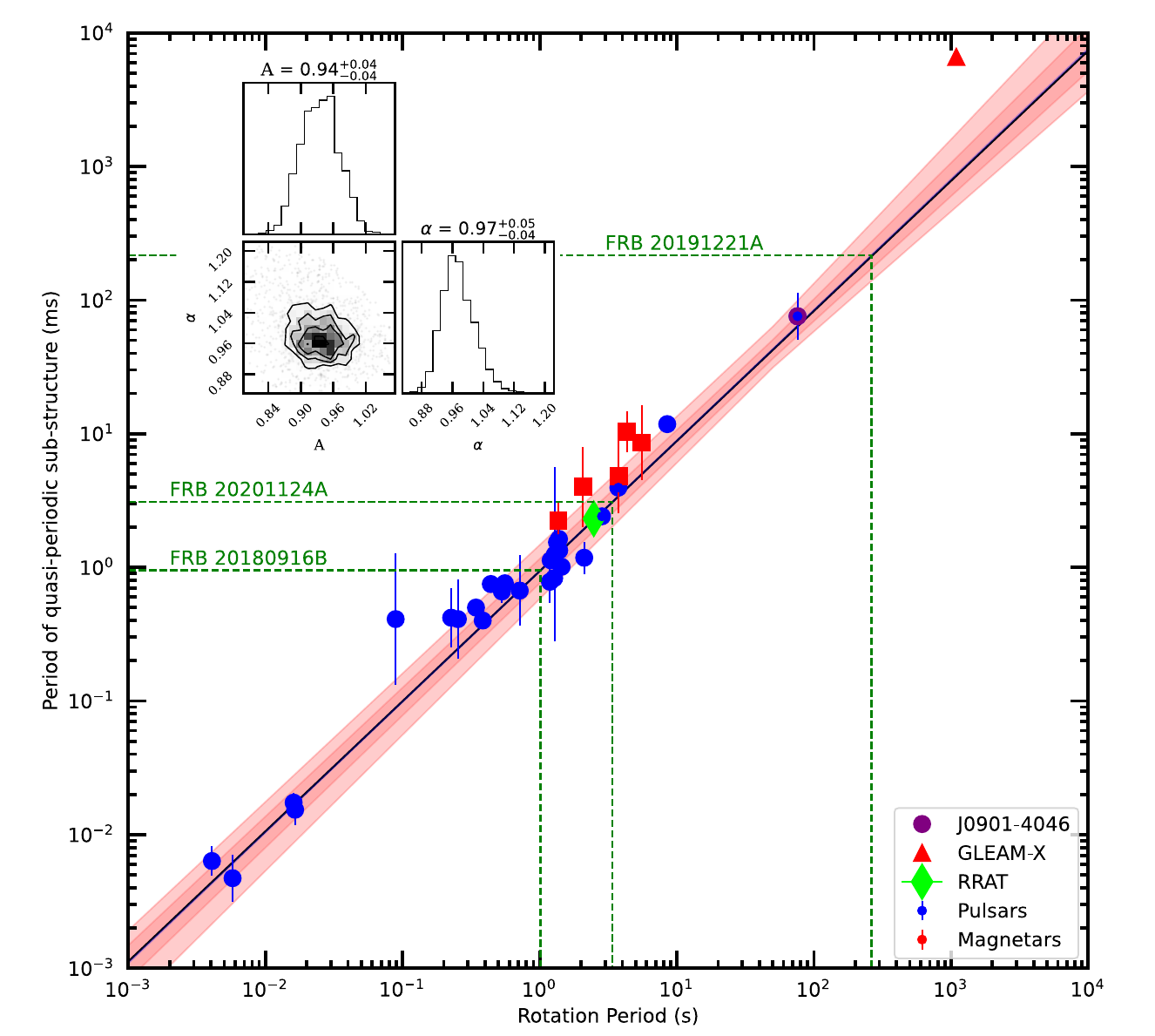}

\caption{Observed relationship of the  quasi-periodicity, $P_\mu$, in the
observed sub-structure
as a function of neutron star
rotation period.
Normal pulsars and millisecond pulsars are shown as blue circles. 
Magnetars studied here are shown as red squares.  
The 76-s pulsar, PSR\,J0901$-$4046, is marked as a purple circle. 
GLEAM-X J162759.5$-$523504.3 is marked with a red triangle. 
{\new 
Each data point represents
the geometrical mean with errorbars determined by the geometric standard
deviation  (see Methods).}
A power-law, $P_\mu = A \times P^\alpha$,
has been fitted to the data, giving a linear relationship, i.e. 
$P_\mu = (0.94\pm0.04) \times P$(s)$^{(+0.97\pm 0.05)}$ ms. 
{\new  The top-left
inset shows a corner plot of the posterior
distributions of the joint model parameters, with the off-diagonal
elements representing the correlations between parameters, and the
diagonal elements denoting the marginalised histograms.
The $\sim1\sigma$ range of uncertainty in the obtained power-law is indicated as a shaded red band.}
We overplot the solution with the value 
of a quasi-periodicity identified for the bursts of a Rotating Radio Transient RRAT J1918$-$0449
as a lime-coloured thin diamond, which also follows the relationship exactly.
Interpreting quasi-periodicities in FRBs as originating from a similar phenomenon,
one can use the observed values to infer 
underlying rotational periods associated with a potential neutron star origin as previously proposed by Refs.~\cite{mpp+21,lab+22,pvb+22}. To demonstrate this, we select a number of FRB quasi-periodicities that were reported as being significant ($>3\sigma$) in Refs.~\cite{mpp+21,abb+22,nzz+22}.}
 
\label{fig:correlation}
\end{figure}

\begin{figure}[!htp]
\centering
\includegraphics[width=1.05\textwidth]{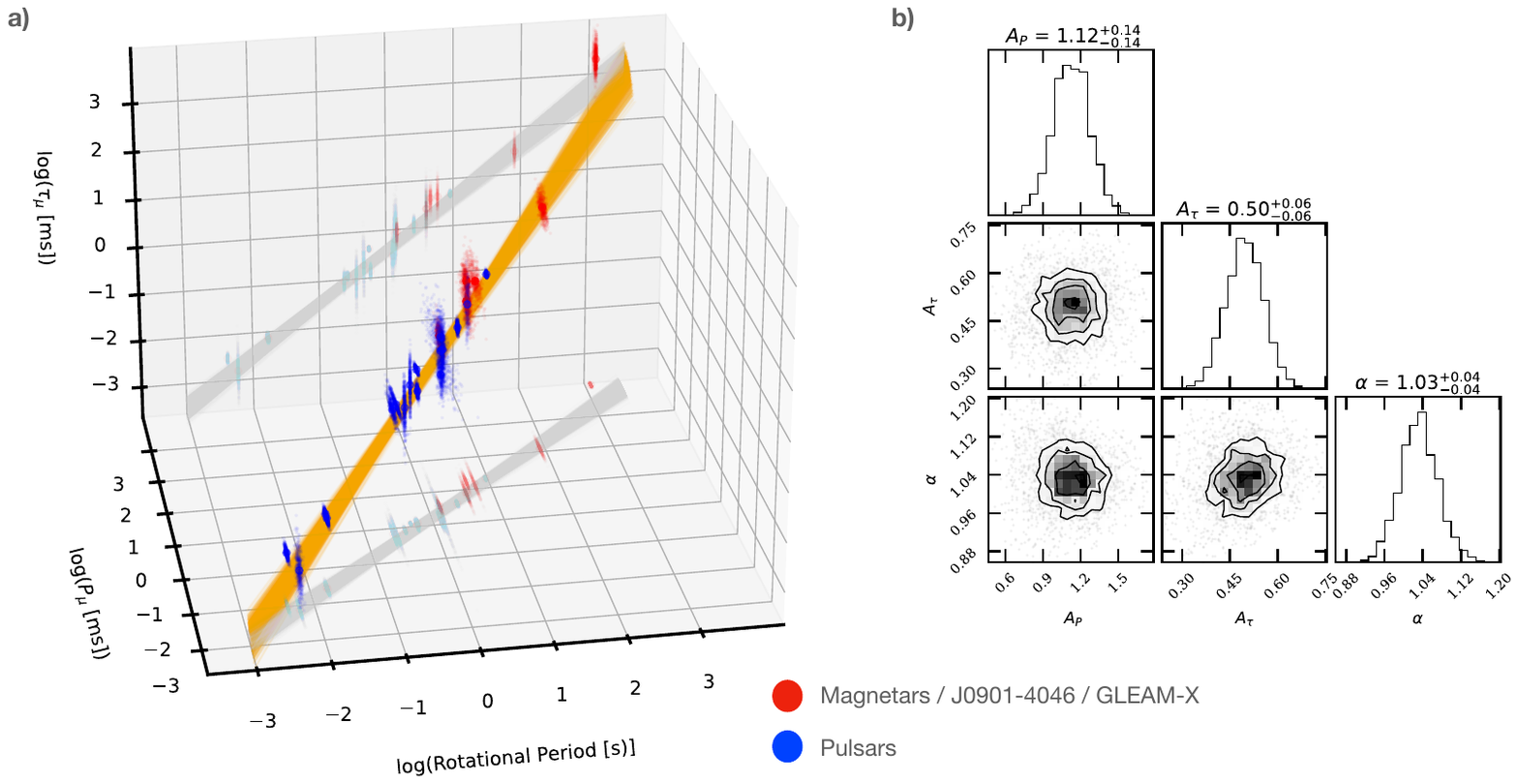}

\caption{
{\new Results of a simultaneous fit of sub-structure periodicity and width to the
rotational period.}
Describing the quasi-periodicity,
$P_\mu$(ms)$= A_P P$(s)$^{\alpha}$,
and width, $\tau_\mu$(ms)$ = A_\tau P$(s)$^{\alpha}$ as a power law with a joint
index $\alpha$, {\new 
the result is shown in panel a) with the studied
radio-loud magnetars, PSR\,J0901$-$4046, and GLEAM$-$X J162759.5$-$523504.3
marked in red. 
The values obtained for pulsars obtained
from the literature (see Methods) are marked in blue. Each data point represents
the geometrical mean surrounded by a cloud with a size determined by the geometric standard
deviation  (see Methods).}
The shaded $\sim1\sigma$ {\new uncertainty-band of solutions is
shown in yellow } with projections on the corresponding planes. 
Panel b) shows a corner plot of the posterior
distributions of the joint model parameters, with the off-diagonal
elements representing the correlations between parameters, and the
diagonal elements denoting the marginalised histograms. We find
$A_P = 1.12\pm 0.14$, $A_\tau = 0.50 \pm 0.06$ and $\alpha = 1.03 \pm 0.04$. }

\label{fig:correlation3d}
\end{figure}

\clearpage 


\clearpage

\section*{Extended Data Figures}


\begin{figure}[!htp]
\centering
\includegraphics[width=0.9\textwidth]{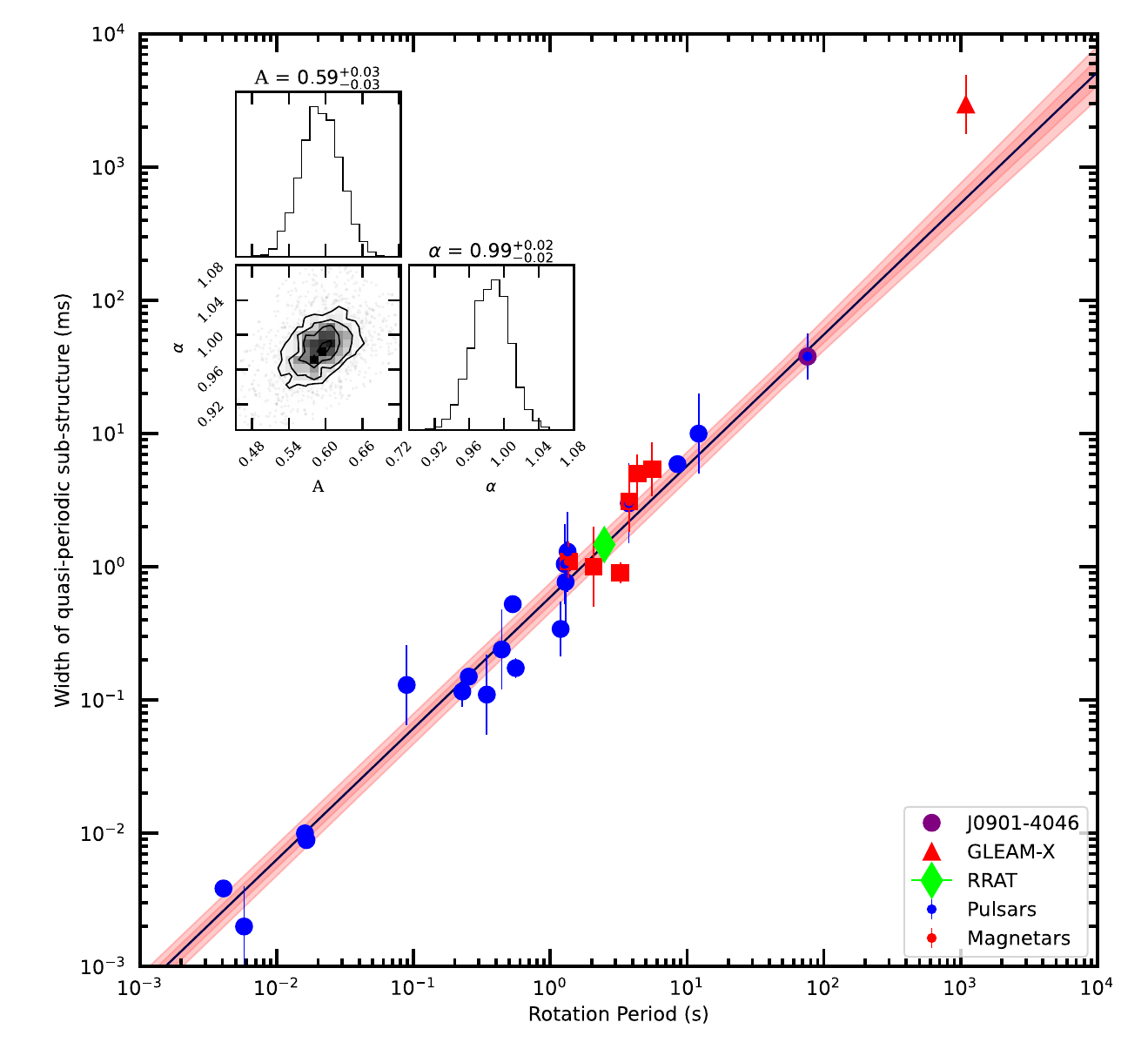}
\caption{Observed relationship of sub-structure
width, $\tau_\mu$, as a function of neutron star
rotation period.
Normal pulsars and millisecond pulsars are shown as blue circles. 
Magnetars studied here are shown as red squares.  
The 76-s PSR~J0901$-$4046 considered to be possibly an old magnetar 
is marked as a red-blue circle. 
GLEAM-X J162759.5$-$523504.3 is marked with a red triangle.
{\new The values obtained for pulsars obtained
from the literature (see Methods) are marked in blue. Each data point represents
the geometrical mean surrounded by a cloud with a size determined by the geometric standard
deviation  (see Methods).}
A power-law, $\tau_\mu = A \times P^\alpha$,
has been fitted to the shown data, giving a linear relationship, i.e. 
$\tau_\mu = (0.59\pm0.03) \times P^{(+0.99\pm 0.02)}$ ms.
{\new  The top-left
inset shows a corner plot of the posterior
distributions of the model parameters, with the off-diagonal
elements representing the correlations between the parameters, and the
diagonal elements denoting the marginalised histograms.
The $\sim1\sigma$ range of uncertainty in the obtained power-law is indicated as a shaded red band.}
A similar
linear dependency on rotation period is found for the quasi-periodicities. 
We also
show the width of quasi-periodic sub-structure identified for the bursts of a Rotating Radio Transient RRAT J1918$-$0449
as a lime-coloured thin diamond, which also fits in the relationship exactly. 
}

\label{fig:correlation2}
\end{figure}

\clearpage

\begin{figure*}[!htp]
\includegraphics[width=1.\textwidth]{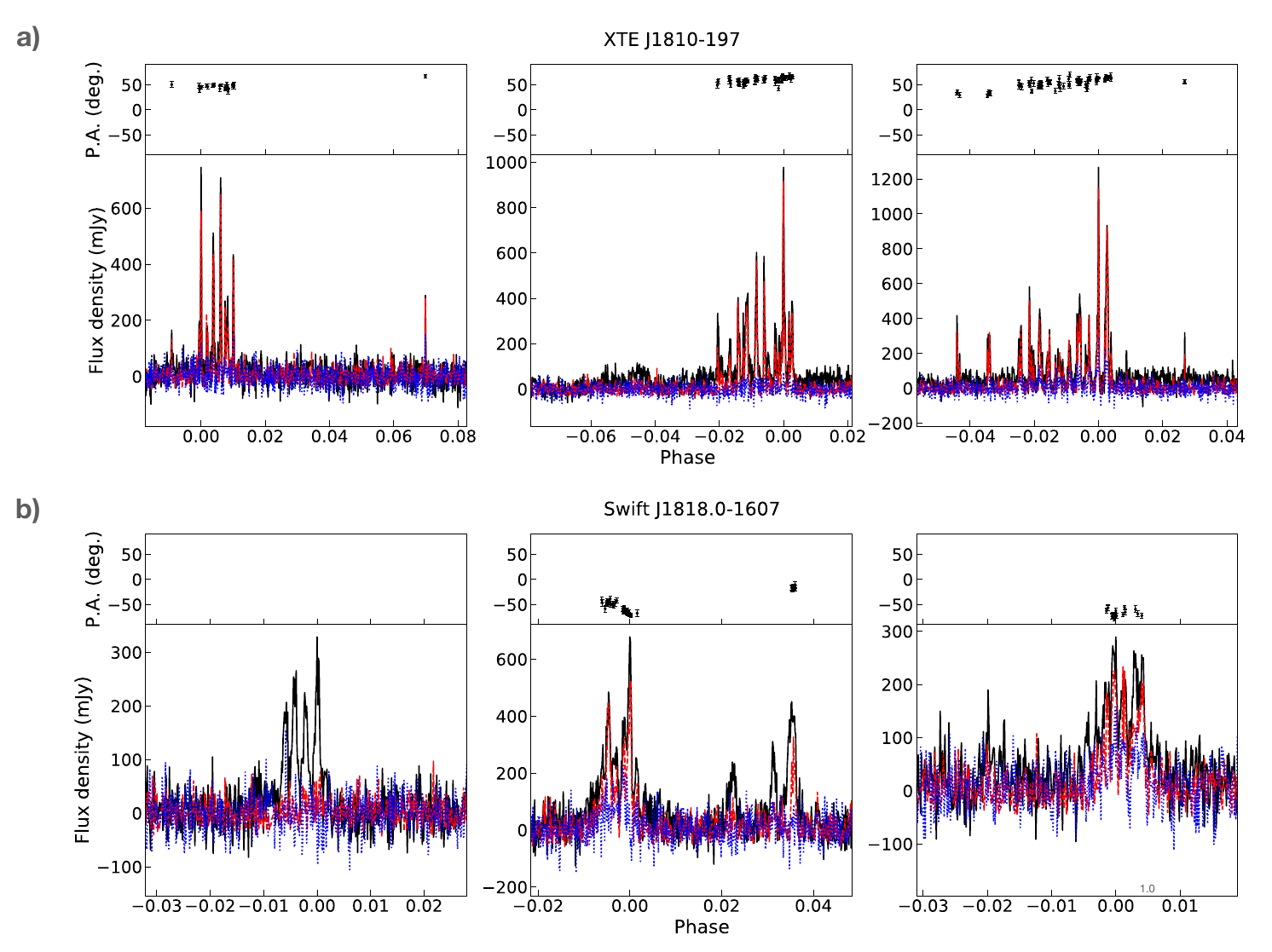}
\caption{Polarisation profile and P.A. of the pulses from XTE~J1810$-$197 and Swift J1818.0$-$1607 shown in Figure~\ref{fig:pulses} (and additional pulses presented in the Supplementary Information.
The black solid, red dashed and blue dotted lines represent total intensity, linear and circular polarisation, respectively.  The pulse from Swift J1818.0$-$1607 in the bottom left panel has no significant linear polarisation. }
\label{fig:polpulse}
\end{figure*}

\clearpage 

\section*{Extended Data Tables}


\begin{table*}[!htp]
\caption{Details of the data analysed for this work. The columns provide the source name, the date of the observation, the central observing frequency ($f$), bandwidth ($\Delta f$), integration time ($t_{\rm obs}$) and time resolution ($\Delta t$) of the data recorded. Data on XTE J1810$-$197 and Swift J1818.0$-$1607 were collected with the Effelsberg Radio Telescope, and those on 1E 1547.0$-$5408 and PSR J1622$-$4950 were recorded at the Parkes Radio Telescope. Note that the observation of GLEAM$-$X J162759.5$-$523504.3 were conducted with the Murchison Widefield Array using its drift scan observing mode, so the integration time is not specified. }
\label{tab:obs}
\footnotesize

\medskip

\begin{tabular}{lccccc}
\hline
\hline
Source & Date & $f$ (GHz)  & $\Delta f$ (MHz) & $t_{\rm obs}$ (min) & $\delta t$ (ms)\\
\hline
1E 1547.0$-$5408  & 2009-02-25 & 8.3 & 512 & 30 & 1.0 \\
PSR J1622$-$4950 
 & 2017-07-04 & 3.1 & 1024 & 15 & 1.0 \\
 XTE J1810$-$197 &2020-08-10   & 6.0       & 4000  & 15 & 0.68\\
   &2020-08-28   & 6.0       & 4000  & 14  & 0.68 \\
  &2020-09-25   & 6.0       & 4000  & 13  & 0.68 \\
Swift J1818.0$-$1607 &2021-02-12   & 6.0       & 4000  & 39  & 0.17\\
    &2021-07-05   & 6.0       & 4000  & 25  & 0.17\\
     &2021-09-16   & 6.0       & 4000  & 30  & 0.17 \\
GLEAM$-$X J162759.5$-$523504.3 & 2018-01-09   & 0.152       & 159  & -  & 500 \\
\hline
\end{tabular}
\end{table*}

\clearpage

\begin{table*}[!htp]
\caption{\small  
{ \new
Measured properties of the  quasi-periodic
sub-structure. For the studied magnetars and  GLEAM$-$X J162759.5$-$523504.3 (abbreviated below as GLEAM$-$X)
we present the width, $\tau_\mu$, and quasi-periodicity, $P_\mu$.}
For both quantities, we quote the mean (MN), the median (MD), the geometric mean (GM) and the geometric standard deviation
factor (GSTD).  For  PSR J1745$-$2900 and SGR J1935$+$2154 the values are displayed
in brackets to indicate that these were derived from the literature, see text for details.}
\label{tab:results}

\medskip

 \setlength{\tabcolsep}{2pt}
\tiny
\begin{tabular}{lc|cccc|cccc}
\hline
\hline
Source & $P$ & \multicolumn{4}{c}{ ----- $\tau_\mu$ ----- } & \multicolumn{4}{c}{ ----- $P_\mu$ ----- } \\
 &  (s) & MN (ms) & MD (ms) & GM (ms) & GSTD & MN (ms) & MD (ms) & GM (ms) & GSTD \\
\hline
1E 1547.0$-$5408 &  2.07 &  -- & -- & 1.0 & 2.0 & -- & -- & 4.0 & 2.0 \\
J1622$-$4950 & 4.33 & $5.3 \pm0.1$ & 4.7 & 5.0 & 1.4 & $11.0 \pm 0.2$ & 10.0 & 10.3 & 1.4 \\
J1745$-$2900 & 3.77 & -- & -- & (3.1) & (1.7) & -- & -- & (4.8) & (1.9) \\
XTE J1810$-$197    & 5.54 &  $6.0\pm0.4$ & 6.8 & 5.4 & 1.6  & $10.3\pm0.6$  & 12.0 & 8.6 & 1.9   \\ 
Swift J1818.0$-$1607 & 1.37 & $1.16\pm0.02$ & 1.0 & 1.1 & 1.4 & $2.37\pm0.04$ & 2.24 & 2.2 & 1.4   \\
SGR J1935$+$2154 & 3.24 & ($0.89\pm0.04$) & (0.93) & (0.88) & (1.2) & -- & -- & -- & -- \\
GLEAM$-$X  & 1090.8 & $3400\pm1800$ & 3200 & 3000 & 1.7 & $6110 \pm 290 $ & 6110 &  6103 & 1.0 \\
\hline
\end{tabular}
\end{table*}

\clearpage

\begin{table}[!htp]
\caption{
 Published values of sub-structure properties in radio pulsars as compiled from the literature/
 Width, $\tau_\mu$, and quasi-periodicity, $P_\mu$, of quasi-periodic
 sub-structure as determined for pulsars and sources not already listed in Table \ref{tab:results}, as inferred from the literature. We quote the name of the source, the rotation period, $P$, and the
 geometric mean (GM) and the geometric standard deviation factor (GSTD) computed for values obtained from the references listed in the last column. Note that it is often more reliable to measure periodicities rather than widths, resulting in more measurements for the former (see e.g.~Ref.~\cite{mar15}). See text for details.}
 \label{tab:psrvals}

\medskip

\begin{tabular}{lcccccl}
\hline
\hline
Source & $P$ & \multicolumn{2}{c}{---- $\tau_\mu$ ----} & \multicolumn{2}{c}{---- $P_\mu$ -----} & References \\
J2000 & (s) & GM (ms) & GSTD  & GM (ms) & GSTD & \\
\hline
0304$+$1932  & 1.3876      & -- & --             &   1.34 &  1.2    & \cite{mar15} \\
0332$+$5434 &  0.7145      & -- & --             &   0.67  &  1.8    & \cite{kkn+78,lkwj98}\\ 
0437$-$4715 &   0.0058    & 0.002 & 2.0       &   0.005 &  1.5  & \cite{dgs16} \\ 
0528$+$2200 &  3.7455      & 3.05 & 2.0      &     3.95 &  1.2    & \cite{cor79,mar15} \\
0546$+$2441 & 2.8439       & -- & --             &      2.41 &  1.1  & \cite{mar15} \\
0659$+$1414 &  0.3849      & -- & --             &   0.40 &  1.1    & \cite{mar15} \\
0814$+$7429 &  1.2922       & 0.77 & 2.0     &  1.25  & 4.5      &   \cite{cwh90} \\
0826$+$2637 &  0.5307      & 0.53 & 1.1     &    0.66 &  1.2    &\cite{cor79,lkwj98,mar15} \\
0835$-$4510 &   0.0893      & 0.13 & 2.0     &   0.41 &  3.1    &\cite{kjv02} \\ 
0837$+$0610  & 1.2738      & 1.05 & 2.0     &      0.83  & 1.2  &\cite{mar15} \\
0901$-$4046 &  75.886       & 37.9 & 1.5          &  75.7  & 1.5     & \cite{Caleb2022} \\
0953$+$0755 &  0.253 1     & 0.15 & 1.1   &    0.41 & 2.0    &\cite{hb81,cwh90,lkwj98, pbc+02b,mar15} \\
1022$+$1001 &  0.0165  & 0.009 & 1.1     &  0.015  &  1.3  &\cite{lab+22} \\
1136$+$1551  & 1.1879      & 0.34 & 1.6     &    0.78 & 1.4       &\cite{fs78,bor83,cwh90,pbc+02b,mar15} \\
1239$+$2453  & 1.3824       & -- & --            &  1.63 & 1.4         &\cite{mar15} \\
1744$-$1134  &  0.0041    & 0.004 & 1.1    &  0.006 &  1.3  &\cite{lab+22}\\
1918$-$0449 &  2.479         & 1.47 & 1.2         & 2.31 & 1.1      & \cite{cwy+22} \\
1921$+$2153  & 1.3373       & 1.30 & 2.0     &      1.54 &  1.6   &\cite{cor79,mar15} \\
1932$+$1059 &   0.2265     & 0.12 & 1.3      &    0.42 &  1.7   & \cite{lkwj98,pbc+02b,mar15} \\
1946$+$1805 &   0.4406     & 0.24 & 2.0      &    0.75 &  1.1    &\cite{cor79,cwh90,mar15} \\
2004$+$3137 &   2.1112     & -- & --             &      1.18  & 1.3  & \cite{mar15} \\
2018$+$2839 &  0.5579      & 0.17 & 1.2     &   0.76 & 1.2    &\cite{cor76a,cor79,cwh90,lkwj98,mar15} \\
2022$+$2854 &   0.3434     & 0.11 & 2.0       &   0.50 &  1.1    & \cite{cor79,mar15} \\
2113$+$2754 &   1.2028     & -- & -               &  1.13 &  1.1       & \cite{mar15} \\
2144$-$3933 &   8.5098      & 5.90  & 1.1        &   11.8 & 1.1     & \cite{mbma20} \\
2145$-$0750 &   0.0161   & 0.011 & 1.1        &   0.017 &  1.2  &\cite{dgs16} \\
2251$-$3711 &  12.123          & 10.0 & 2.0      & -- & --             & \cite{mke+20}\\ 
2317$+$2149 &  1.4447      & -- & --             & 1.01  & 1.1       & \cite{mar15} \\
\hline
\end{tabular}

\end{table}

\newpage

\section*{Supplementary Information}

\setcounter{figure}{0}
\setcounter{table}{0}

\renewcommand{\thepage}{S\arabic{page}} 
\renewcommand{\thesection}{S\arabic{section}}  
\renewcommand{\thetable}{S\arabic{table}}  
\renewcommand{\thefigure}{S\arabic{figure}}

\label{sec:supplement}

\subsection*{Statistical tests on quasi-periodic signals}

\begin{figure}[!htp]
\centering
\includegraphics[scale=0.6]{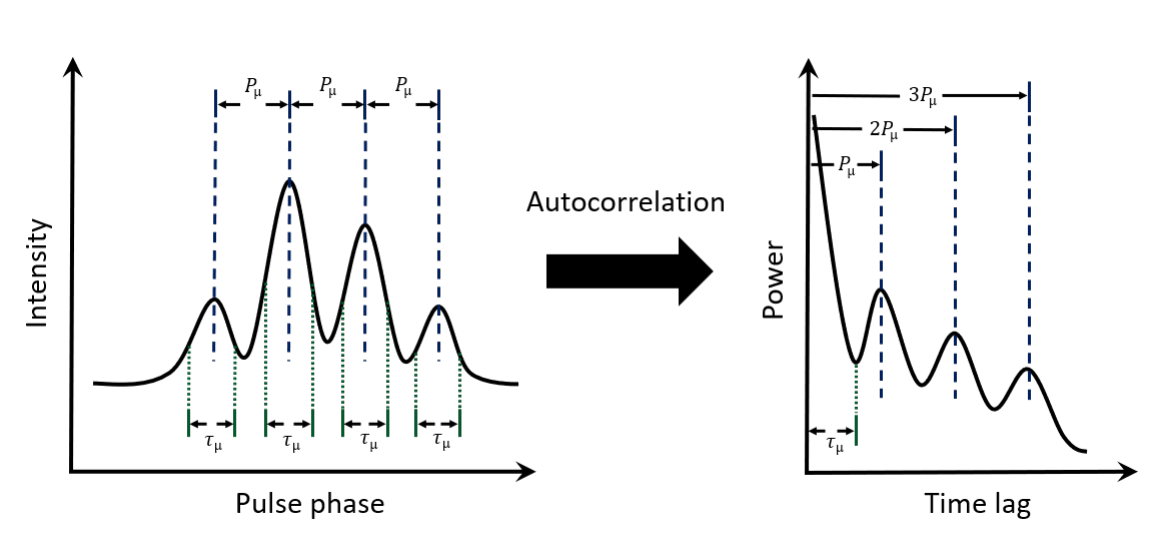}
\caption{ Illustration of the first principle for measuring the quasi-periodicity ($P_{\rm \mu}$) and width ($\tau_{\rm \mu}$) of sub-structure
using the auto-correlation analysis.}
\label{fig:acf_sketch}
\end{figure}

\begin{figure}[!htp]
\centering
\includegraphics[width=0.9\textwidth]{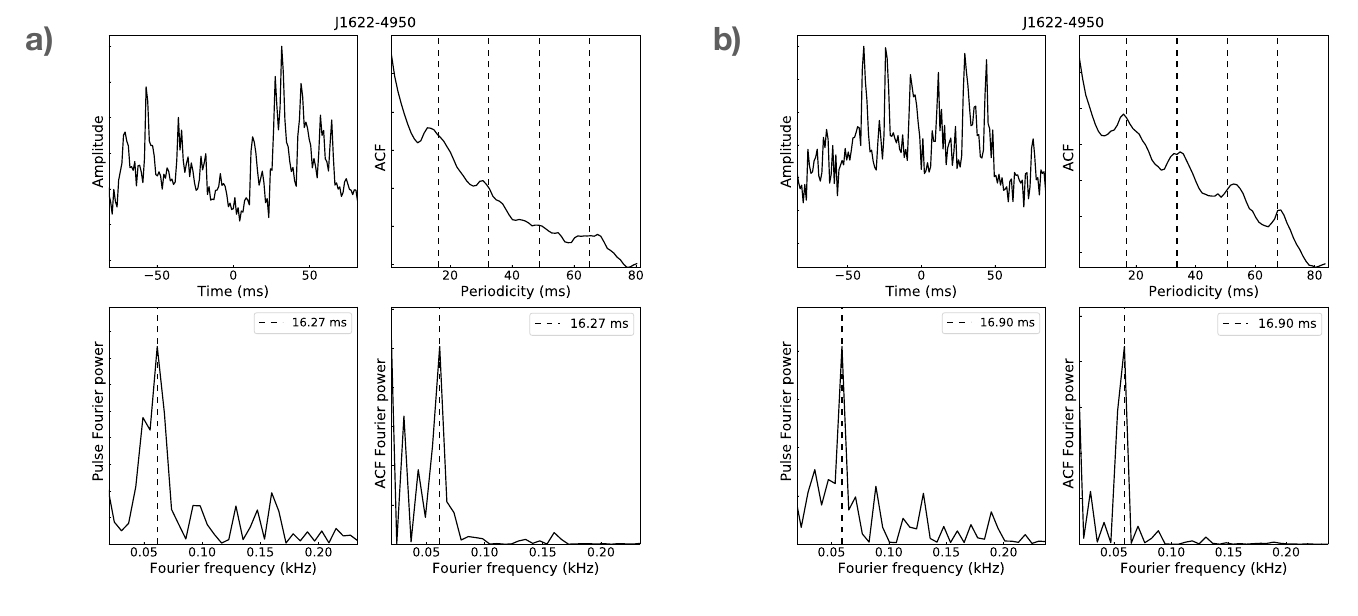}
\caption{Two more examples of quasi-periodic sub-structure 
from J1622$-$4950. 
The four sub-panels show pulse amplitude as a function of time, the auto-correlation (ACF) function of the pulse, the Fourier power spectral density (PSD) of the pulse, and the PSD of the ACF, respectively. The vertical dashed lines mark the identified quasi-periodicity and its harmonics in the ACF panels, and their corresponding fundamental frequency in the PSD panels.
}

\label{fig:1622pulse}
\end{figure}

\begin{figure}[!htp]
\centering
\includegraphics[width=0.9\textwidth]{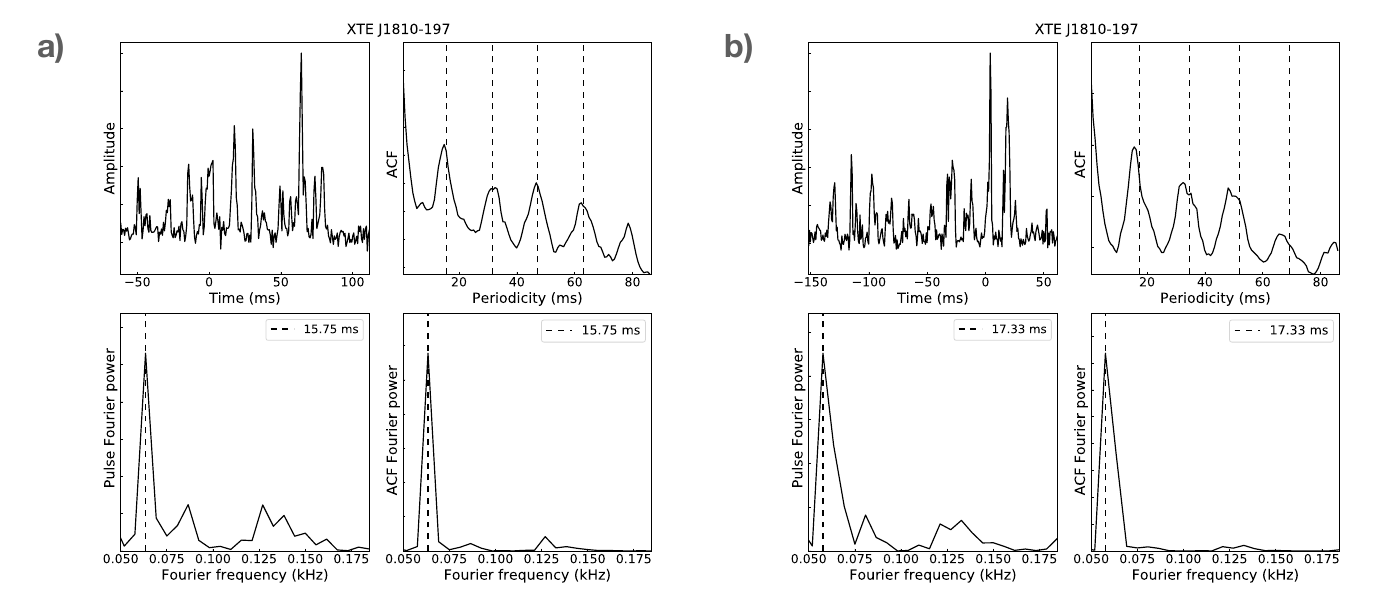}
\caption{Two more examples of quasi-periodic sub-structure
from XTE~J1810$-$197. T
The four sub-panels show pulse amplitude as a function of time, the auto-correlation (ACF) function of the pulse, the Fourier power spectral density (PSD) of the pulse, and the PSD of the ACF, respectively. The vertical dashed lines mark the identified quasi-periodicity and its harmonics in the ACF panels, and their corresponding fundamental frequency in the PSD panels.
}
\label{fig:1810pulse}
\end{figure}

\begin{figure}[!htp]
\centering
\includegraphics[width=0.9\textwidth]{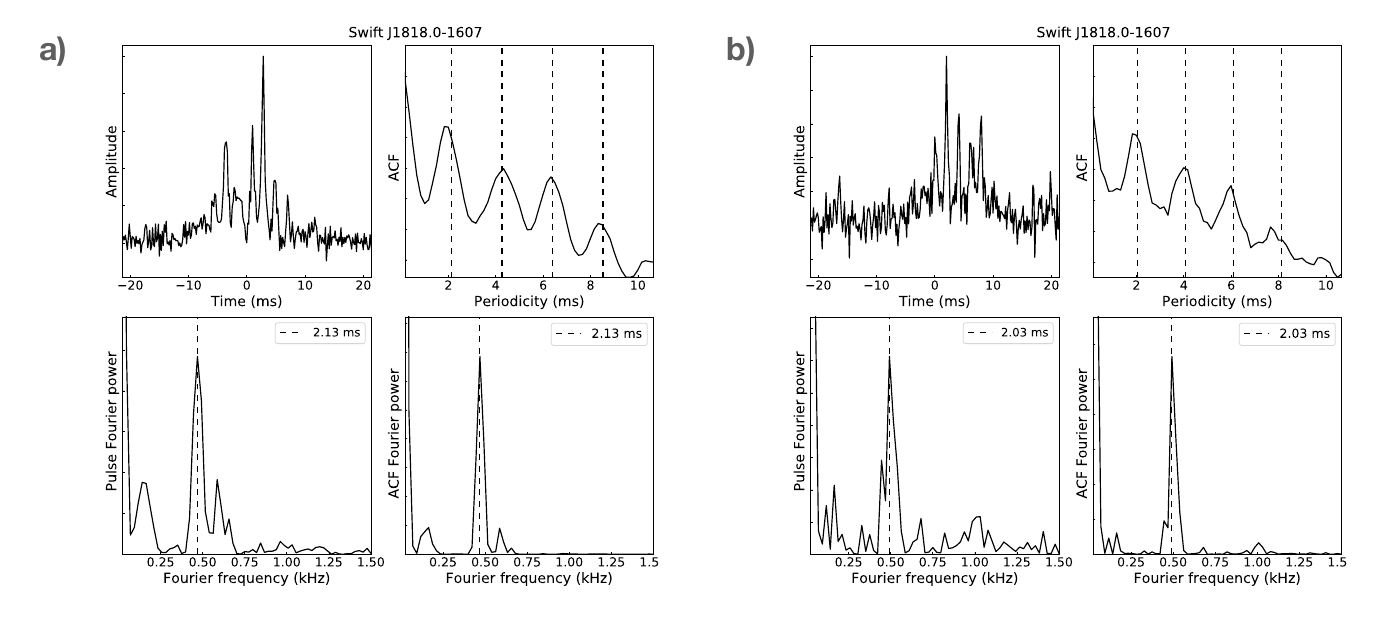}
\caption{Two more examples of quasi-periodic sub-structure
from Swift J1818.0$-$1607. 
The four sub-panels show pulse amplitude as a function of time, the auto-correlation (ACF) function of the pulse, the Fourier power spectral density (PSD) of the pulse, and the PSD of the ACF, respectively. The vertical dashed lines mark the identified quasi-periodicity and its harmonics in the ACF panels, and their corresponding fundamental frequency in the PSD panels.}
\label{fig:1818pulse}
\end{figure}

\begin{figure}[!htp]
\centering
\includegraphics[width=0.9\textwidth]{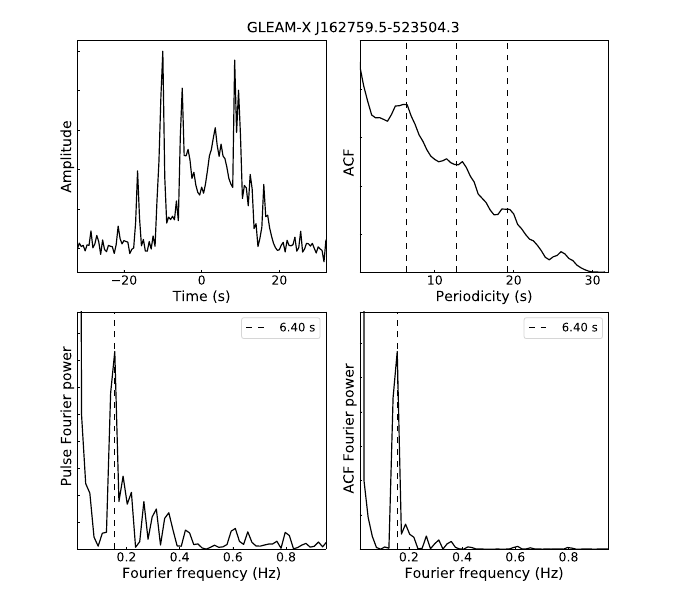}
\caption{One more example of quasi-periodic sub-structure from GLEAM$-$X J162759.5$-$523504.3. 
The four sub-panels show pulse amplitude as a function of time, the auto-correlation (ACF) function of the pulse, the Fourier power spectral density (PSD) of the pulse, and the PSD of the ACF, respectively. The vertical dashed lines mark the identified quasi-periodicity and its harmonics in the ACF panels, and their corresponding fundamental frequency in the PSD panels.}
\label{fig:GLEAMpulse}
\end{figure}

{\new The method to determine sub-pulse structure and its properties as described in 
the Methods section is illustrated in Figure~\ref{fig:acf_sketch}. Further examples
of applying this method are shown in Figures~\ref{fig:1622pulse}, \ref{fig:1810pulse},
\ref{fig:1818pulse} and \ref{fig:GLEAMpulse}. }

\begin{figure}[!htp]
\centering
\includegraphics[width=0.95\textwidth]{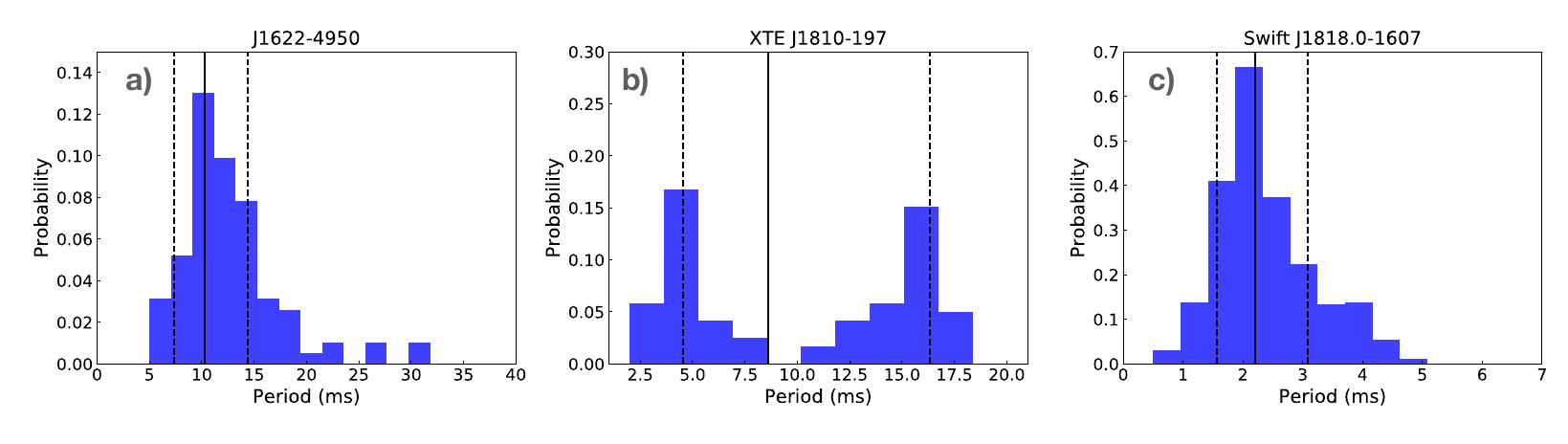}
\caption{
Histogram of quasi-periodicities measured in Swift J1818.0$-$1607, XTE J1810$-$197, PSR J1622$-$4950. {\new  In each panel, the solid line denotes geometrical mean of the distribution. The range of
uncertainty is determined from the
geometric standard deviation (GSTD) factor, $\Delta x_{GSTD}$, as from
$x_{\rm geo}/\Delta x_{GSTD}$ to $x_{\rm geo} \times \Delta x_{GSTD}$,
which is indicated in each plot by the dashed lines.} }
\label{fig:microP}
\end{figure}

In order to evaluate the significance of those detected quasi-periodicities,
we performed and compared a
number of statistical tests. In the following, we describe those tests. In addition to two tests closely
related to our detection method described in detail earlier, we also studied the suitability of the
Rayleigh test, which is a standard method to detect and quantify potential periodicity in data 
using the Rayleigh ($Z^2_1$) statistic. Recently, this latter method has been applied frequently 
in the evaluation of possible quasi-periodicities in FRBs. 
The various methods are in particular useful, if only single or few bursts are available, as in the case of most FRBs.
Obviously, the ability of studying large samples of pulses (as available for normal pulsar observations and
demonstrated in many past studies, or for 
radio loud magnetars studied here) can provide additional and even stronger evidence by building up robust statistics
from many independent measurements, leading to histograms like Figure~\ref{fig:microP}.

\bigskip

\begin{figure*}
\centering
\includegraphics[width=0.95\textwidth]{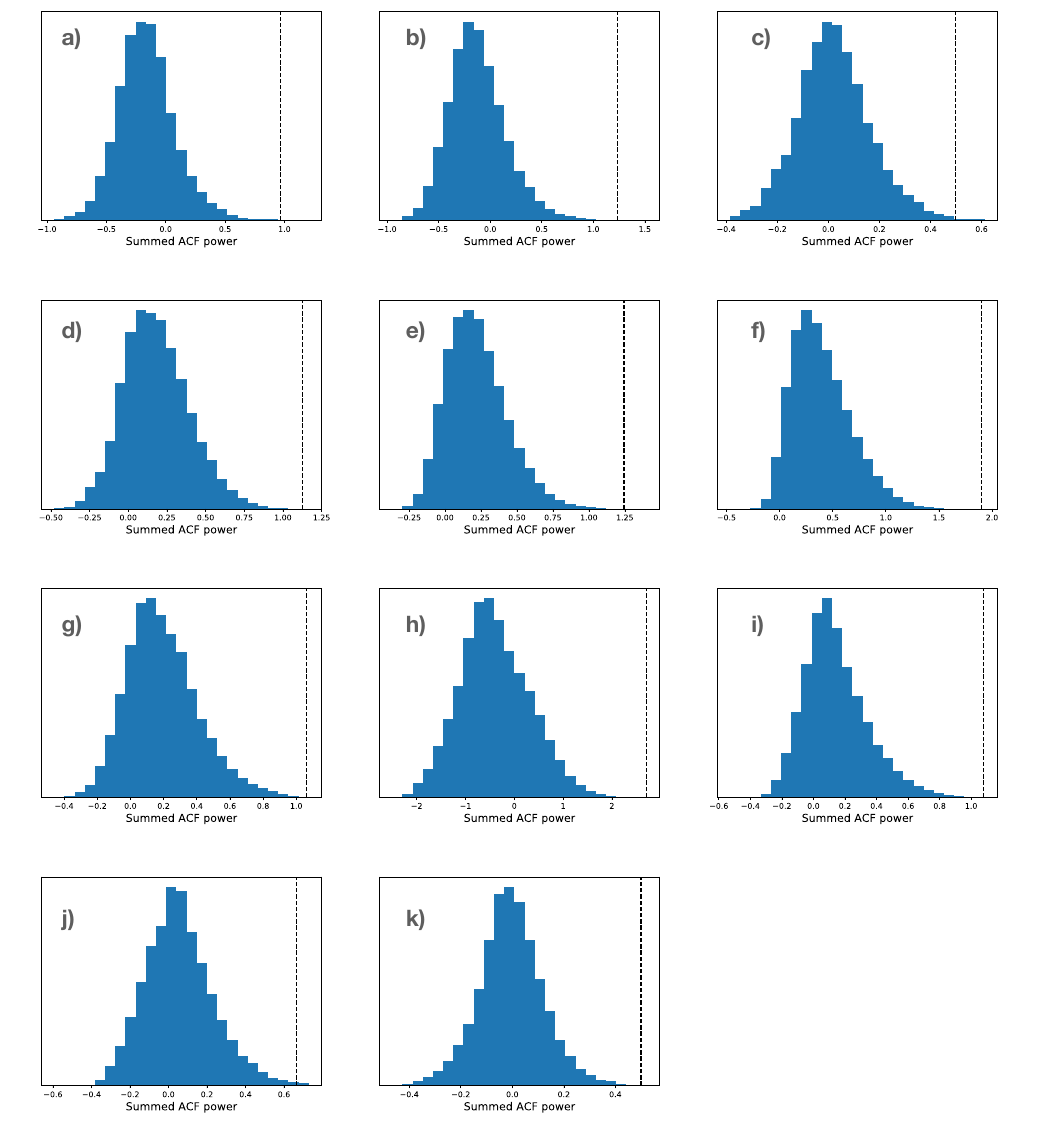}

\caption{Summed ACF power compared to the null distribution for all sample pulses shown in this paper. 
{\new
Panels a), b) and c) give the results for PSR~J1622$-$4950 from Figure 1 and Figure~\ref{fig:1622pulse} (left and right, respectively). 
Panels d), e) and f) give the results for XTE J1810$-$197 from Figure 1 and Figure~\ref{fig:1810pulse} (left and right, respectively). 
Panels g), h) and i) give the results for Swift J1818.0$-$1607 from Figure 1 and Figure~\ref{fig:1818pulse} (left and right, respectively). 
Panels i) and k) give the results for GLEAM$-$X J162759.5$-$523504.3 from Figure 1 and Figure~\ref{fig:GLEAMpulse}, respectively. 
In each panel, the dashed line marks the summed ACF power measured from the real pulse.} The corresponding detection significances are presented in Table~\ref{tab:zscore}. \label{fig:ACF_CPS_dist}}
\end{figure*}

\noindent
{\bf ACF-based test: component phase scrambling.} \label{ssec:comp_scram_test}
This test is applied to all example pulses shown in Table~\ref{tab:zscore}. 
 As discussed in the Methods Section, the periodic / quasi-periodic features in the data will result in a sequence of local maxima in the ACF with spacing equalling to the periodicity. Thus, one can construct a statistical quantity $\mathscr{P}$, by summing the power of these relative maxima in the ACF ($\rho$) that correspond to the periodicity:
\begin{equation}
    \mathscr{P}=\sum_i \rho(P_{\rm \mu}\cdot i),~~P_{\rm \mu}\cdot i < l/2,
\end{equation}
where $l$ is the length of the data. Thus, the detection significance can be obtained by comparing this quantity from the real data to a null distribution. To construct the null distribution, we simulated a number of pulses and calculated their corresponding $\mathscr{P}$ values. Each simulated pulse consists of the same number of components as in the example pulse, each created using the measured sub-structure width and amplitudes from the pulse. The centers of these components were computed by adding random variations to a set of equally spaced positions separated by the measured periodicity. The random variation was drawn from a uniform probability distribution defined as Equation~(9) of \cite{abb+22}, with the same scaling value of $\chi=0.2$. Then for all simulated pulses, we obtained the periodicity from the ACF analysis and calculated the corresponding $\mathscr{P}$ values from their ACFs. The detection significance was thus obtained by comparing the $\mathscr{P}$ value from the real pulse to its distribution constructed from these simulated pulses, as shown in Figure~\ref{fig:ACF_CPS_dist}.

The resulting detection significance from this analysis are summarized in Table~\ref{tab:zscore}. It can be seen that for all pulses from XTE J1810$-$197, Swift J1818.0$-$1607 and one from GLEAM$-$X J162759.5$-$523504.3, the significance is higher than 4$\sigma$; this suggests a strong evidence for the presence of periodicity in the data. The rest pulses all reach a significance between 3 and $4\sigma$, indicating the existence of a weaker evidence.

\bigskip

\begin{figure}[!htp]
\centering
\includegraphics[width=0.9\textwidth]{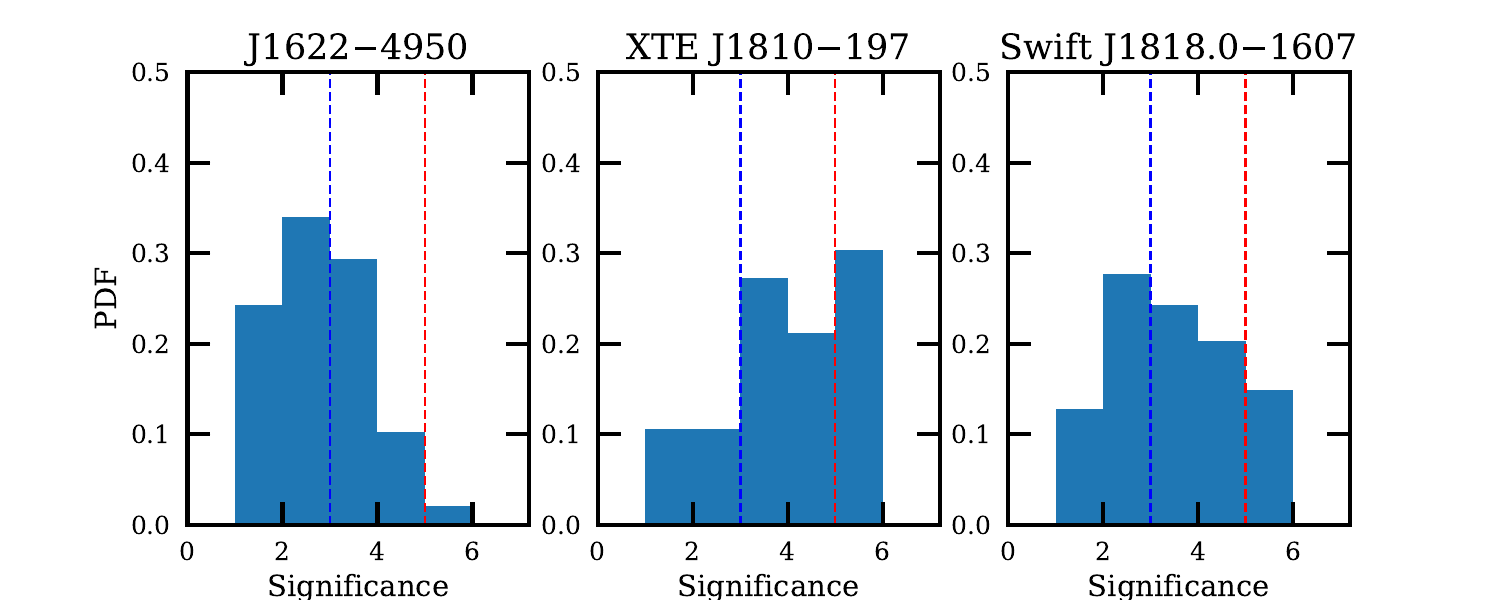}
\caption{Histogram of significances determined for 
quasi-periodicities measured in PSR J1622$-$4950, XTE J1810$-$197 and Swift J1818.0$-$1607. {\new The blue and red lines in each panel correspond to significance values of 3 and (at least) 5$\sigma$, respectively. Most studied
pulses show significant periodicities. See text for details.} }
\label{fig:significance}
\end{figure}

\noindent
{\bf ACF-based test: phase bin scrambling.} \label{ssec:comp_scram_test} This test is similar to the previous one but goes one step further by comparing the observed
pulses with representations of themselves that have the order of the corresponding phase bins
scrambled completely in a random fashion. This method is similar to the recipe provided,
for instance, in Ref.~\cite{vaug05}: For each pulse passing the criteria for containing 
a quasi-periodicity as determined by the method described in the Methods Section,
the phase bins of the pulse window are scrambled in a random fashion. This randomized pulse
is subjected to the same ACF computation as the real pulse, i.e.~the resulting ACF
is Fourier-transformed to calculate the Power Spectral Density. For this, the maximum peak (ignoring
the zeroth and Nyquist frequency bins) is recorded. The maximum peak obtained from the original, unscrambled
pulse is compared to the distribution of recorded peaks from thousands of such random pulses.
The resulting statistics allows to compute the corresponding significance levels. In order
to limit the number of random pulses required to be studied, 
we stopped the computations when a significance of more than
$5\sigma$ was established. Figure \ref{fig:significance} shows the distributions of significance
values as established for all pulses of those three sources, where many observed pulses
were available and studied in detail, i.e.~for Swift J1818.0$-$1607, XTE J1810$-$197, PSR J1622$-$4950. 
In this figure, we mark a threshold of $3\sigma$ and $5\sigma$. The latter was marked separately, since 
the last bin (displayed from $5\sigma$ to $6\sigma$) contains all pulses, where the significance is exceeding
$5\sigma$ or more. The derived significance for the quasi-periodicities seen in all other pulses and sources
displayed in the various figures of this work is summarised in Table~\ref{tab:zscore} for a comparison with the other
methods. We note the significance values are very much consistent among the first two methods but
typically exceed the significance given by the Rayleigh tests for the reasons demonstrated in the following.

\bigskip

\noindent
{\bf Rayleigh test.} Using the Rayleigh statistic, Refs.~\cite{abb+22}, \cite{pvb+22} and \cite{nzz+22} conducted significance tests of periodicity on several FRBs with quasi-regularly spaced features. These returned a high significance ($>3\sigma$) for the presence of periodicity only for three of the overall samples. Here, following a similar recipe as described in \cite{abb+22}, we carried out the Rayleigh statistic test to the example pulses shown in Figure~1, \ref{fig:1810pulse}, \ref{fig:1818pulse}, \ref{fig:1622pulse} and \ref{fig:GLEAMpulse}. In detail, we first smoothed the pulse data using continuous wavelet transformation, and recorded the positions of the relative maxima detected with sufficiently high S/N and when the peaks were at least as wide as two to three phase bins. Then we conducted a blind search for periodicity, by finding the maximum of Rayleigh ($Z^2_1$) test statistic
\begin{equation}
    Z^2_1=\frac{2}{N}\left[\left(\sum^{N}_{i=1}\cos\phi_i\right)^2 + \left(\sum^{N}_{i=1}\sin\phi_i\right)^2\right],
\end{equation}
where $N$ is the number of peaks and $\phi_i$ is the phase of the peak positions with respect to the trial period. The trial periods range from five times the sampling interval up to half the entire duration of the on-pulse region, with a step size one fifth of the sampling interval. The detected periodicity was then the trial period corresponding to the maximum $Z^2_1$ value. 

To estimate the corresponding significance of the $Z^2_1$ value, we followed the {\new recipe} described in \cite{abb+22} to construct the null distribution, by conducting Monte Carlo simulations based on the measured peak positions. In each iteration, we added a random variation to the peak positions drawn from a uniform probability distribution defined as Equation~(9) of \cite{abb+22}. Here we assumed an exclusion parameter of $\chi=0.2$ as in the analysis of \cite{abb+22}. Then a maximum $Z^2_1$ value was obtained by following the same procedure as described above. Eventually, we compared the distribution of the maximum $Z^2_1$ values from $20000$ such iterations, to the measurement from real data. The false-alarm probability was calculated based on the number iterations that have a maximum $Z^2_1$ value larger than the measurement from real data. 

\begin{figure*}
\centering
\includegraphics[width=0.95\textwidth]{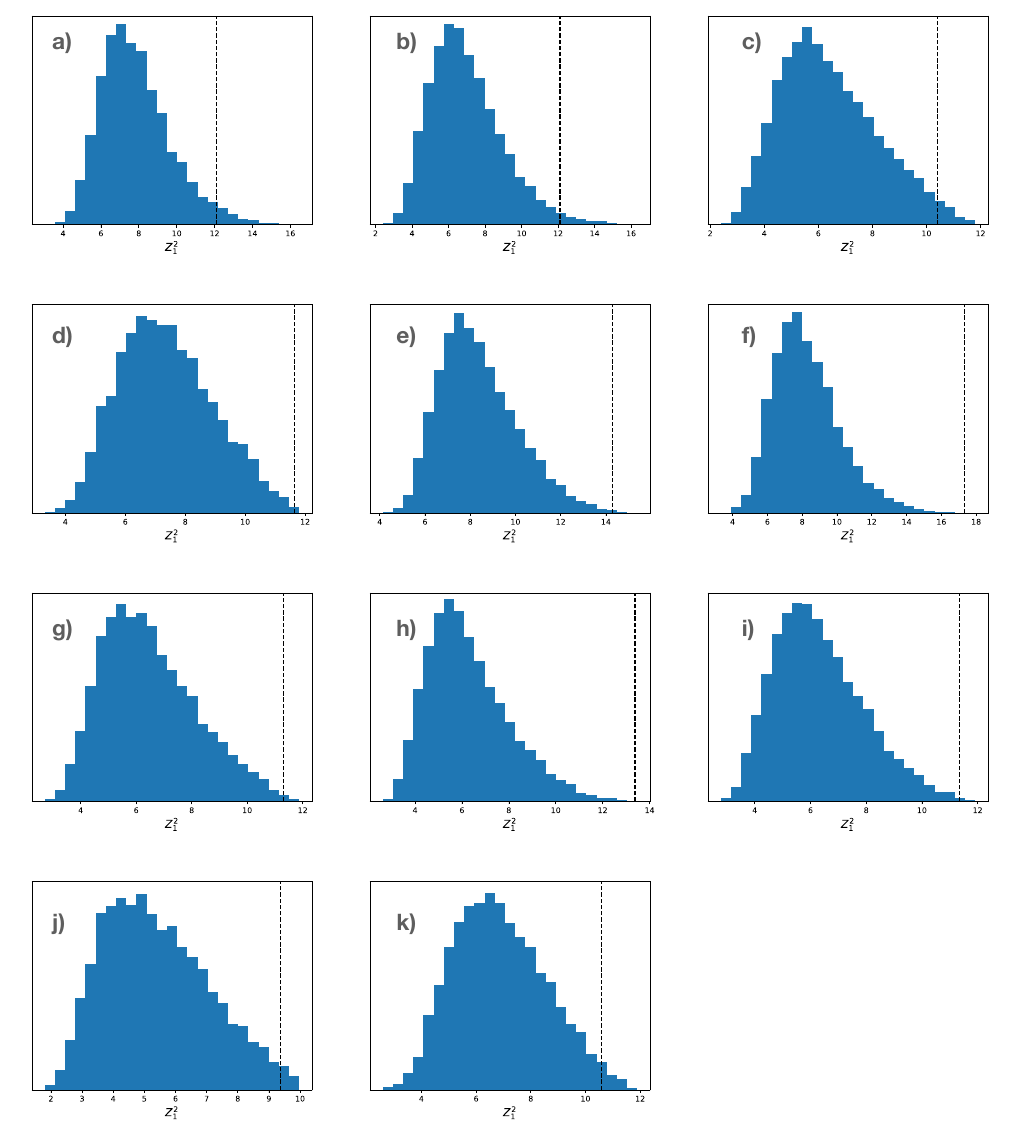}

\caption{ 
\new
Maximum $Z^2_1$ values as determined for all sample pulses shown in this paper. 
Panels a), b) and c) give the results for PSR~J1622$-$4950 from Figure 1 and Figure~\ref{fig:1622pulse} (left and right, respectively). 
Panels d), e) and f) give the results for XTE J1810$-$197 from Figure 1 and Figure~\ref{fig:1810pulse} (left and right, respectively). 
Panels g), h) and i) give the results for Swift J1818.0$-$1607 from Figure 1 and Figure~\ref{fig:1818pulse} (left and right, respectively). 
Panels i) and k) give the results for GLEAM$-$X J162759.5$-$523504.3 from Figure 1 and Figure~\ref{fig:GLEAMpulse}, respectively. 
In each panel, the dashed line marks the maximum $Z^2_1$ value from all pulses, compared to the null distribution of the Rayleigh test constructed from $20000$ iterations. The corresponding detection significances are presented in Table~\ref{tab:zscore}. \label{fig:ztest_dist}}
\end{figure*}

The results for all of our samples are shown in Table~\ref{tab:zscore}, while the obtained $Z^2_1$ values in comparison with the null distribution can be found in Figure~\ref{fig:ztest_dist}. It can be seen that all sample pulses from XTE J1810$-$197 and Swift J1818.0$-$1607 return a significance higher than 3$\sigma$. In all cases of PSR~J1622$-$4950 and GLEAM$-$X J162759.5$-$523504.3 have such a value above 2. 

However, it should be noted that the Rayleigh statistical test is designed and optimal for verifying signals of a \emph{precise} period. Therefore, as we show in the following, the
test can fail to identify {\em quasi}-periodicities in the signal, which is exactly the type of features seen in numerous pulsars by far \cite{cwh90,lkwj98,dgs16,lab+22}. 
In contrast, this type of signal would still manifest itself as a strongly preferred times/frequency scale in the ACF and PSD. To demonstrate this, we have conducted a series of mock data simulations. For each of the four sources, XTE J1810$-$197, Swift J1818.0$-$1607, PSR~J1622$-$4950 and GLEAM$-$X J162759.5$-$523504.3, we created a set of mock pulse profiles based on a template which consists of a sequence of Gaussian components: 
\begin{equation}
    P(t)=\sum_{i=1}^{N_{\rm p}}A_{\rm g}e^{-(t-c_{i})^2/2\sigma_{\rm g}^2}
\end{equation}
Here $N_{\rm p}$ is the number of components. These components are precisely equally spaced in the first place. The values of component spacing and width used in the simulation for each of the source were obtained from the real measurements shown in Table~2. Then in each iteration, we created a new pulse by altering the centre ($c_{i}$) of each component randomly within a range of $(c_{i}-J\cdot\sigma_{\rm g}$, $c_{i}+J\cdot\sigma_{\rm g})$, where $J$ specify the level of phase variability. This pulse-to-pulse signal variability, also known as ``pulse jittering'', is commonly seen in pulsar radio emission \cite[e.g.,][]{cd85}. With this variability being present, the simulated pulsar signal usually shows quasi-regularly spaced pattern instead of a precise periodicity. 

\begin{figure*}
\centering
\includegraphics[width=0.95\textwidth]{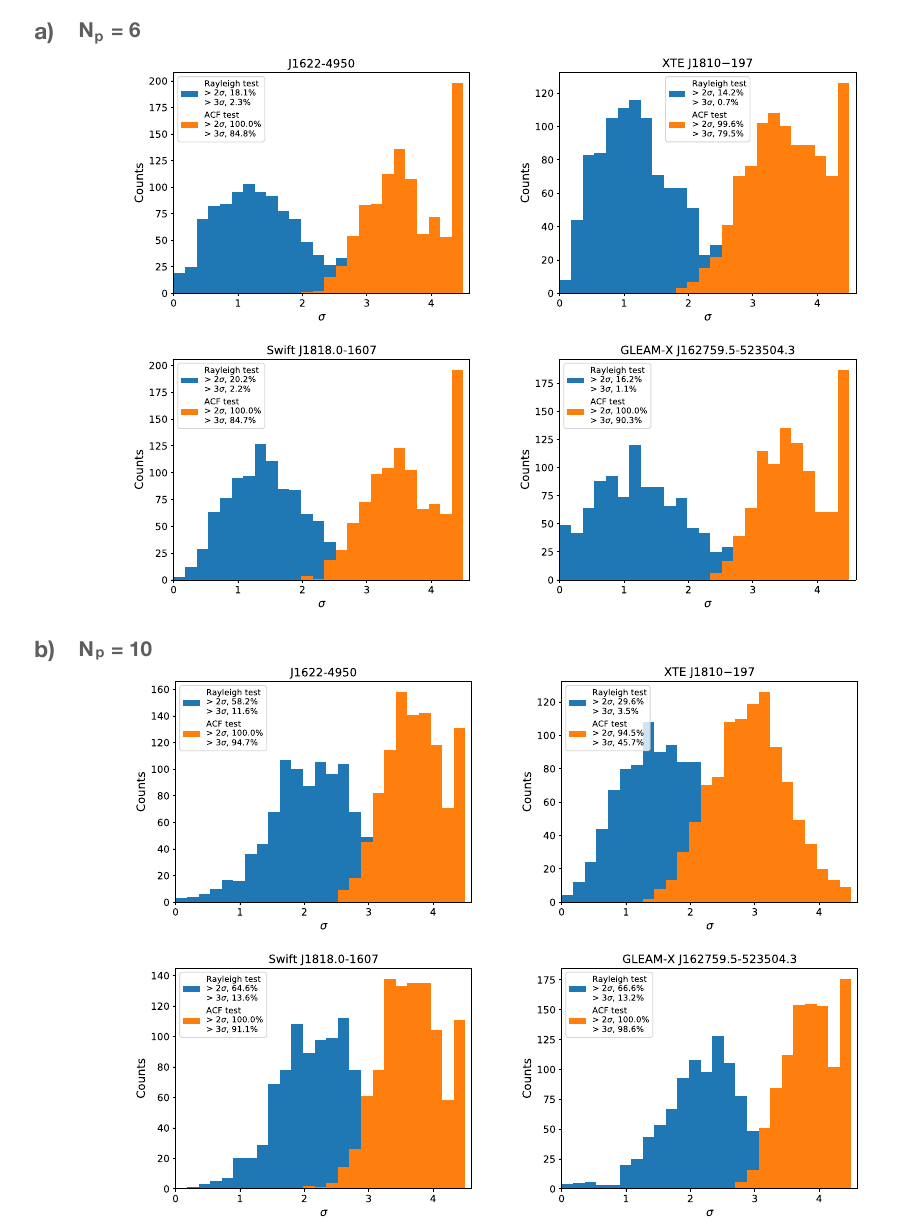}

\caption{Distribution of detection significance ($\sigma$) for the obtained periodicities, from both the Rayleigh test and the ACF test with component phase scrambling. For each source, {\new we conducted two sets of simulations, with a) $N_{\rm p}=6$ and b) $N_{\rm p} = 10$, respectively. In each set}, $10^3$ pulses were simulated using the measured properties in Table~3. Note that in the ACF test, for each pulse we conducted $5\times10^4$ iterations of component phase scrambling, which corresponds to a limit of approximately 4.4$\sigma$. This is the reason for the excess in counts on the right hand side of the histogram. \label{fig:sigma_simu}}
\end{figure*}

\begin{figure*}
\centering
\includegraphics[width=0.95\textwidth]{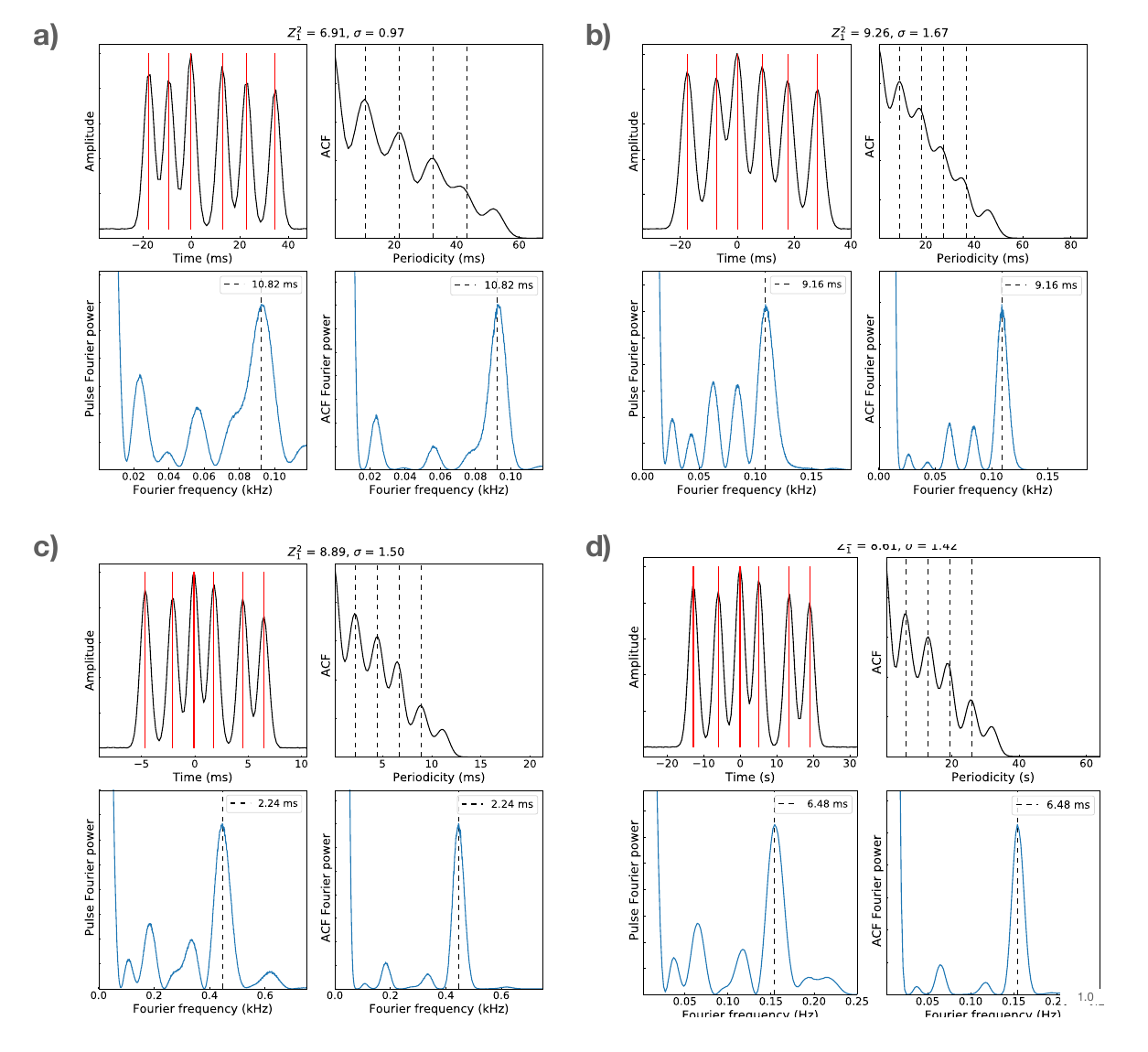}

\caption{Examples of created pulses from the mock data simulations as described in the main texts. 
The corresponding sources are a) PSR~J1622$-$4950, b) XTE J1810$-$197, c) Swift J1818.0$-$1607
and d) GLEAM$-$X J162759.5$-$523504.3. The four sub-panels show pulse amplitude as a function of time, the auto-correlation (ACF) function of the pulse, the Fourier power spectral density (PSD) of the pulse, and the PSD of the ACF, respectively. The vertical dashed lines mark the identified quasi-periodicity and its harmonics in the ACF panels, and their corresponding fundamental frequency in the PSD panels.
The heading of each panel a) - d) provides the results of the  Rayleigh test.}
\label{fig:MCsample}
\end{figure*}

Next for each simulated pulse, we followed the same steps as described above to obtain the detected periodicity, the Rayleigh test statistics and the significance of periodicity. As a comparison, we also obtained the detection significance of periodicities from the phase-bin-scrambling ACF test as described above. The distribution of the significance from $10^3$ iterations are shown in Figure~\ref{fig:sigma_simu} for the case of all four sources. 
{\new Here, we conducted two sets of simulations, one with $N_{\rm }=6$ and the other with $N_{\rm p}=10$, so as to cover the range of number of components seen in our selected samples as shown in Table~\ref{tab:zscore}. In both sets, we used} 
a typical value of $J=1$. Four sample pulses (one for each source) from the simulation 
{\new with $N_{\rm }=6$ are shown in Figure~\ref{fig:MCsample}. It can be seen that  for the simulation set with $N_{\rm }=6$}, 
in all cases only a small fraction, i.e., $<20\%$ of the simulated pulses return a significance of periodicity above $2\sigma$ and no more than 2\% of them are above $3\sigma$ significance. In contrast, the low-significance pulses show strong evidence for quasi-periodicity in the Fourier-domain and ACF analysis as can be seen in Figure~\ref{fig:MCsample}. {\new When the number of components in each simulated pulse is increased from 6 to 10, the performance of Rayleigh test is improved as also anticipated in \cite{abb+22}. Still, only a small fraction ($<14\%$) of simulated pulses yield a significance above $3\sigma$ significance, and the contrast between the distributions of significance obtained from the two methods is still apparent.} Overall, the ACF tests return a significantly higher detection significance of periodicity. These demonstrate that the apparent significance derived from the statistical test can be easily affected when the feature in the signal deviates from being precisely periodic, even though we start from a highly periodic signal. In contrast, the detection scheme based on PSD and ACF as applied routinely in pulsar astronomy,
and also in this work, can still correctly identify the underlying quasi-periods and widths. 

\begin{figure*}
\centering
\includegraphics[scale=0.5]{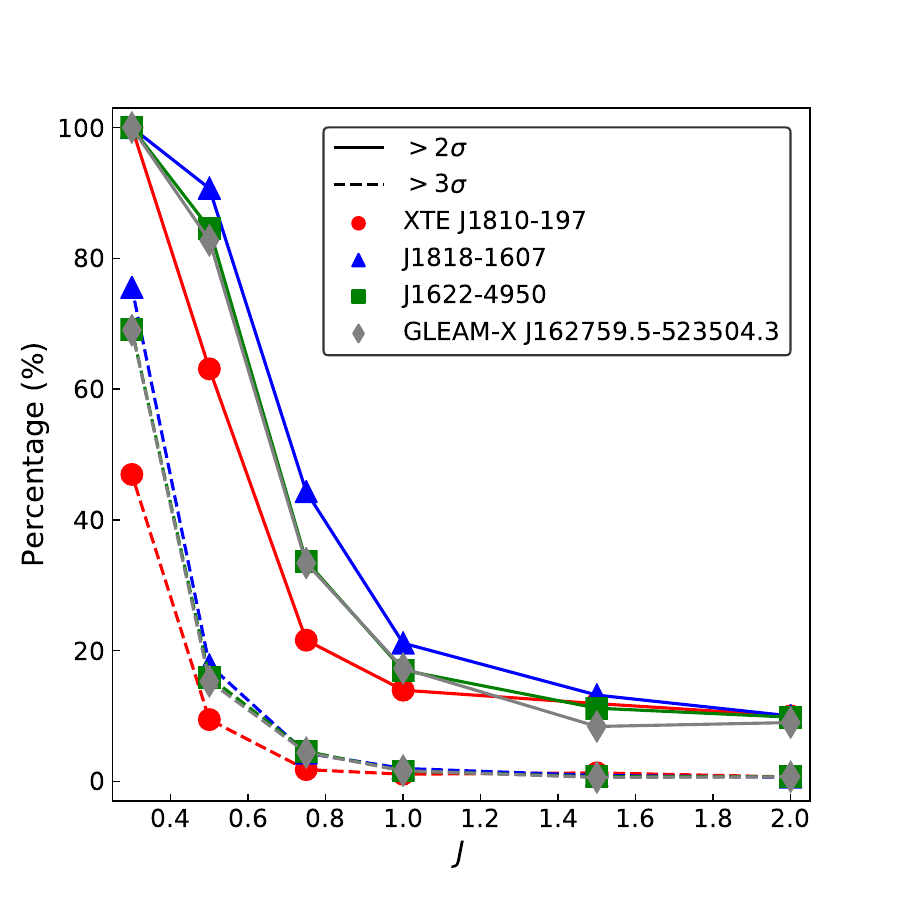}
\caption{Performance of the Rayleigh test for pulses with different intensity of phase variability. Here the solid, dashed lines represent the percentage of in total $10^4$ simulations that return a detection significance higher than 2 and 3$\sigma$, respectively. The results of all four sources presented in Table~\ref{tab:zscore} are shown. \label{fig:sigma_simu_J}}
\end{figure*}

To further investigate the impact by pulse phase variability on the performance of Rayleigh test, we extended the simulation study above to a range of values for $J$. Then for each source and choice of $J$, we again simulated $10^4$ pulses and recorded the percentage of these iterations that returned a detection significance above 2 and 3$\sigma$ from the Rayleigh test. From Figure~\ref{fig:sigma_simu_J}, it can be seen that for $J>1$, the Rayleigh test is of barely any sensitivity to the embedded quasi-periodicity. When the intensity of phase variability decreases, i.e., the periodicity becomes more precise, the output by Rayleigh test then has a higher probability to return a high detection significance. For $J\lesssim0.3$, all iterations return a detection significance higher than 2 and a large fraction higher than 3, which is the case for all four sources studied in this paper.

In summary, the performance of Rayleigh test depends strongly on the intrinsic pulse phase variability when it is used for evaluating the presence of quasi-periodicity. For
this reason, we prefer the established method applied in pulsar astronomy over the past decades, to study the quasi-periodic feature in the data.


\clearpage

\clearpage

\begin{table*}
\caption{{\new Summary of significance tests for shown pulses.}
The number of peaks ($N_{\rm p}$), reported periodicity ($P_{\rm \mu, ACF}$, $P_{\rm \mu, Z}$) and detection significance ($\sigma_{\rm CPS}$, $\sigma_{\rm PBS}$, $\sigma_{\rm Z}$) from the and the Rayleigh tests, for the example pulses shown in Figure~1, \ref{fig:1810pulse}, \ref{fig:1818pulse}, \ref{fig:1622pulse} and \ref{fig:GLEAMpulse}. Here the detection significance from both the component phase scrambling (CPS) and phase bin scrambling (PBS) ACF tests are shown. \label{tab:zscore}}
\footnotesize
\begin{tabular}{lccccccl}
\hline
Source & $N_{\rm p}$ & $P_{\rm \mu, ACF}$ (ms) & $\sigma_{\rm CPS}$ & $\sigma_{\rm PBS}$ & $P_{\rm \mu, Z}$ (ms) & $\sigma_{\rm Z}$ & Reference \\
\hline
XTE J1810$-$197 &6 & 11.6 & 4.1 & $>5$ & 11.2 & 3.2 & Figure~1 \\
& 8 & 15.8 & 4.1 & $>5$ & 15.6 & 3.2 & Figure~\ref{fig:1810pulse} (left) \\
& 10 & 17.3 & 4.9 & $>5$ & 16.9 & 3.7 & Figure~\ref{fig:1810pulse} (right)\\
Swift J1818.0$-$1607 & 6 & 2.84 & 4.1 & $>5$ & 2.78 & 3.0 & Figure~1 \\
& 7 & 2.13 & 4.6 & 4.9 & 2.13 & 3.9 & Figure~\ref{fig:1818pulse} (left) \\
& 6 & 2.03 & 4.4 & 4.7 & 2.03 & 3.2 & Figure~\ref{fig:1818pulse} (right) \\
PSR~J1622$-$4950 & 9 & 16.7 & 3.8 & $>5$ & 16.6 & 2.4 & Figure~1 \\
& 9 & 16.3 & 3.6 & 4.4 & 16.8 & 2.3 & Figure~\ref{fig:1622pulse} (left) \\
& 6 & 16.9 & 3.0 & 4.3 & 17.2 & 2.3 & Figure~\ref{fig:1622pulse} (right)\\
GLEAM$-$X & 5 & 5824 & 3.2 & $>5$ & 5925 & 2.5 & Figure~1 \\
J162759.5$-$523504.3 & 6 & 6404 & 4.2 & 4.4 & 6711 & 2.4 & Figure~\ref{fig:GLEAMpulse} \\
\hline
\end{tabular}
\end{table*}

\end{document}